\documentstyle[aps,prb,preprint,epsf]{revtex}

\begin{document}
\title{Frictional drag between quantum wells mediated 
by phonon exchange} 

\preprint{IUCM97-018}

\author{Martin Chr. B{\o}nsager$^{1}$, Karsten Flensberg$^{2}$,
Ben Yu-Kuang Hu$^{3}$, and A.H. MacDonald$^{1}$}

\address{$\mbox{}^1$ Department of Physics, Indiana University,
Bloomington, Indiana 47405}

\address{$\mbox{}^2$ Danish Institute of Fundamental Metrology,
Bldg. 307, Anker Engelunds Vej 1,
DK-2800 Lyngby, Denmark}

\address{$\mbox{}^3$ Mikroelektronik Centret,
Bldg.\ 345east, Technical University of Denmark,
DK-2800 Lyngby, Denmark}

\date{\today}
\maketitle

\begin{abstract}

We use the Kubo formalism to evaluate the contribution of
acoustic phonon exchange to the frictional drag 
between nearby two-dimensional electron systems. 
In the case of free phonons, we find a divergent 
drag rate ($\tau_{D}^{-1}$).  However, $\tau_{D}^{-1}$ 
becomes finite when phonon scattering from 
either lattice imperfections or electronic excitations
is accounted for. 
In the case of GaAs quantum wells, we find that for a phonon
mean free path $\ell_{\rm ph}$ smaller than a critical value,
imperfection scattering dominates and 
the drag rate varies as $\ln (\ell_{\rm ph}/d)$ over 
many orders of magnitude of the layer separation $d$.  
When $\ell_{\rm ph}$ exceeds the critical value, the drag rate is 
dominated by coupling through an electron-phonon collective mode
localized in the vicinity of the electron layers. 
We argue that the coupled electron-phonon mode may be observable 
for realistic parameters.
Our theory is in good agreement with experimental results for the 
temperature, density, and $d$-dependence of the drag rate.
\end{abstract}


\section{Introduction}

Interactions between particles are a cornerstone of much of
today's research in physics.  In nuclear and high-energy physics,
the effects of these interactions can be probed directly through
scattering experiments.  In condensed matter physics, 
inter-particle interaction effects are enriched by the close proximity of 
other particles giving rise to a plethora of fascinating phenomena.  
However, direct measurement of these interactions in a condensed 
matter system is often a more difficult exercise, because of 
the indirect way in which scattering amplitudes are 
related to observables. 

Some time ago, Pogrebinskii and later Price\cite{first} proposed  
the following direct probe of interparticle interactions through
a transport measurement.  Place two 2-dimensional (2D) electron films close
enough together and draw a current in one film.  Through
interlayer interactions, net momentum is transferred to electrons
in the adjacent film, inducing a current there which can be
measured.  Due to technological difficulties in contacting the
individual layers, decades passed before the first frictional drag 
experiment between 2D and 
3-dimensional (3D) layers was performed.\cite{solo89}
The first experiments on this phenomenon between 
two 2D systems, as originally envisaged in Ref.~\onlinecite{first} 
were performed by Gramila {\em et al.} for two electron 
layers,\cite{gram-coul,gram-phon} and by Sivan {\em et al.} for an 
electron--hole system.\cite{siva92} 
In these experiments a current is drawn in the first layer,
while the second layer is an open circuit.  Instead of a current 
in the second layer there will be an induced electric field that 
opposes the ``dragging force''
from the first layer. The {\em transresistivity} $\tensor\rho_{21}$ 
is defined as the ratio of the induced electric field in the
second layer to the driving current density in the first, 
\begin{equation}
\tensor\rho_{21}\cdot{\bf J}_{1} = {\bf E}_{2}. 
\end{equation}
The stronger the interlayer interaction, the larger the
magnitude of the transresistivity.  (In this paper, we shall
treat isotropic systems at zero magnetic field,
hence $\tensor\rho_{21}$ is diagonal). 
The transresistivity is often interpreted in terms of a drag rate 
which, in analogy with a Drude model, is defined by 
$\tau_D^{-1}=\rho_{21}n_1e^2/m^*$ where $n_1$ is the electron density of 
the driving layer and $m^*$ is the electron effective mass.

These experiments spurred a large body of theoretical work 
both on electron-hole systems \cite{elechole} and on electron--electron
systems.\cite{eetheo,mosko,tso92,zhan93,jauh93,zhen93,kame95,flen94,flen95a,flen95b,ssgrecent}
Most of this work has focused on interlayer 
Coulomb interaction, the most obvious coupling mechanism and 
the one considered in the original theoretical papers.\cite{first}
However, it was clear from the start that the experimental results
were inconsistent with a purely Coulomb interlayer interaction,
which predicted a low temperature\cite{gram-coul,jauh93,zhen93} 
($k_{B}T \ll \varepsilon_{F,1},\varepsilon_{F,2}$, 
where $\varepsilon_{F,i}$ is the Fermi energy for layer $i$)
transresistivity of the form
\begin{equation}\label{coul}
\rho_{21}=\left(-\frac{h}{e^2}\right)\frac{\zeta(3)\pi}{32}
\frac{(k_{B}T)^2}{\varepsilon_{F,1}\varepsilon_{F,2}}\frac{1}
{(k_{F,1}d)\,(k_{F,2}d)}\,\frac{1}{(q_{\rm TF}d)\,(q_{\rm TF}d)},
\end{equation}
where $\zeta$ is the Riemann zeta function, $d$ is the interlayer
separation, $k_{F,i}$ is the Fermi wavevector for layer $i$, 
and $q_{\rm TF}$ is the Thomas Fermi screening wavevector of the 2D electron gas.
This expression is based on the random phase approximation
(RPA) for the screened interlayer Coulomb interaction and applies for 
$q_{\rm TF} d \gg 1$ and $k_{F,i} d \gg 1$.
{}From inspection of Eq.\ (\ref{coul}), one notes 
three important characteristics of the Coulomb drag for 
low temperatures:
(1) the scaled transresistivity $\rho_{21}(T)/T^2$ is a constant;
(2) $\rho_{21}$ is a monotonically
decreasing function of the density of either layer 
(so long as $k_{F,i} d \gg 1$); and
(3) $\rho_{21} \propto d^{-4}$.
The experimental results at around 2---3~K, on the other 
hand, showed\cite{gram-coul,gram-phon,rube95} 
(1) a well-defined peak in the $\rho_{21}(T)/T^2$,
(2) a local maximum in $\rho_{21}$ near equal layer densities
and (3) an approximately $d$-independent residual $\rho_{21}$,
after subtraction of the Coulomb contribution which can be identified
experimentally by its simple $d$ and $T$ dependencies.
Furthermore, the experimentally measured magnitude of $\rho_{21}$ 
was generally larger by about a factor of 2 than the value predicted 
by the Coulomb interaction alone.  
Another momentum transfer mechanism was clearly involved. 

{}From the outset it was understood that exchange of acoustic phonons was the 
most likely candidate for this second momentum transfer mechanism.
Exchange of phonons often dominates 
electron-electron scattering contributions to the resistance of
bulk metals.\cite{bulkphononmediated}  
The peak in the temperature dependence of $\rho_{21}(T)/T^2$ 
is reminiscent of features in the temperature dependence of the 
acoustic phonon limited mobility, and occurs
around the Bloch-Gr\"{u}neisen temperature $T_{\rm BG} = 2 k_{B}^{-1}
\hbar c_{l} k_{F}$ associated with the acoustic phonon modes.
($c_{l}$ is the longitudinal acoustic phonon velocity.)
The phase-space available for scattering is largest for $2k_{F}$ transfers,
partially explaining the enhanced drag when $k_{F,1} = k_{F,2}$.
Finally, the long-ranged exchange of phonons between electrically 
isolated systems is not an unknown phenomenon.  Exchange of phonons 
between two 3D systems separated by 
$\sim 100\,\mu$m has been observed previously,\cite{hubn60} and
related effects are expected to be observable in a superlattice if 
the driving electron layer is hot\cite{hot}.
The theoretical challenge is to explain the magnitude of the 
observed drag and its dependence on layer separation and density.

Since interactions of acoustic phonons with electrons are 
relatively weak in GaAs, the fact that phonon-mediated and 
Coulomb contribution to the drag
are often comparable seems mysterious.  However, we show below
that the obvious calculation, in which a free phonon propagator 
substitutes for the Coulomb interaction, leads to a 
divergent drag resistivity.  The large but finite drag 
rates which are observed experimentally can be explained in 
terms of scattering and interaction effects which 
alter the phonon propagator.

Despite the apparent importance of phonon exchange in 
drag measurements,  
there has been less theoretical work on this mechanism 
than on the Coulomb coupling mechanism.  
An incoherent phonon exchange model studied by 
Gramila {\em et al.} produced a drag rate which was 
too weak to account for the observed transresistivity.\cite{gram-phon}
Tso {\em et al.}\cite{tso92} addressed the question of whether 
exchange of {\em virtual} phonons could give a contribution strong enough to
explain the magnitude of the transresistivity, but did 
not use an electron-phonon coupling model which is realistic
for $\rm GaAs/Al_{x}Ga_{1-x}As$ systems.  
Zhang and Takahashi\cite{zhan93} included all relevant phonon
contributions; however, due to an incorrect effective two-dimensional
electron--phonon interaction, they predicted a short range
phonon-mediated drag.
In spite of the use of radically different models for the 
phonon-mediated process, both these calculations yielded a
temperature dependence for the drag 
in reasonable agreement with experiment.
The similarity is not surprising because of the common
appearance of the  Bloch-Gr\"{u}neisen temperature scale
associated with the acoustic phonon mode.  Comparable  
features in the drag, associated with plasmon modes of the 
electronic system, were predicted\cite{flen94,flen95b} and
observed\cite{hill97,gram97} at higher temperatures.

In this paper we report on a detailed examination of the 
phonon exchange mechanism for drag, using a model which 
we believe to be quantitatively reliable for GaAs/AlAs 
quantum well systems.  We 
address distance-, temperature-, and  
density dependence of the transresistivity. 
We show that two different regimes of layer 
separation dependence can occur, depending on  
the phonon mean free path and the electron--phonon coupling
constant. We demonstrate the existence of a coupled electron-phonon
mode, which for long phonon mean free paths leads to a large enhancement
of the drag.

The outline of the paper is as follows. In Section~\ref{formalism} 
we use a Kubo linear response formalism to obtain a formula for the 
transresistivity which is sufficiently general to 
permit the incorporation of a
finite phonon mean-free-path and renormalization of
the phonon propagator due to coupling to the electronic layers.  
In Section~\ref{inter} 
we discuss the phonon mediated {\em e--e} interaction and explain 
how the relatively weak electron--phonon interaction can lead to
a surprisingly large contribution to the drag. The two different 
regimes of $\ell_{\rm ph}$ are discussed in Section~\ref{couplemode} 
and \ref{phdamped}, where approximate expressions for $\rho_{21}$ are 
derived and special attention is paid to the layer separation dependence 
of the drag. Detailed numerical results are presented in 
Section~\ref{num} before we conclude in Section~\ref{conclusion} with 
a summary of results. 

\section{Formalism}\label{formalism}

Identical theoretical expressions for the drag rate due to Coulomb
interactions been obtained in several different ways.  The most 
physically transparent derivation is based on semiclassical
transport theory.\cite{jauh93}
More elaborate fully quantum mechanical derivations
based on memory function\cite{zhen93}
or Kubo formula approaches\cite{kame95,flen95a} yield identical
results for the large $k_{F} \ell$ limit at zero magnetic field
but are more flexible and, in particular, can be applied 
in the presence of an external magnetic field.   
(Here $\ell$ is the electronic mean free path.) 
We show below that at lowest nonvanishing 
order in the electron--phonon interaction, phonon exchange 
yields an infinite result for the drag rate.
The Kubo formula approach, which we use in this paper, 
is most convenient when interlayer interactions 
need to be treated beyond leading order, because 
of the powerful diagrammatic perturbation theory expansion available
to evaluate the influence of interaction terms on the 
appropriate current-current 
correlation function.  The required calculation is an adaptation of 
those described in Refs.~\onlinecite{kame95} and 
\onlinecite{flen95a} and is outlined below.

\subsection{Electron-phonon interaction Hamiltonian}

We consider frictional drag between two GaAs quantum wells.
We define the plane of the quantum wells as the $x$--$y$ plane.
The distance between the centers of the two quantum wells is $d$,
and the width of the two wells is $L$. 
We assume that the electron number densities in each well,
$n_{1}$ and $n_{2}$, are such that only one subband of the quantum
well is occupied.  The formalism can easily be generalized to
accommodate more occupied subbands.

We include interlayer and intralayer Coulombic electron-electron
interactions and the coupling of electrons in either layer  
to the three-dimensional (3D) phonons of the semiconductor
host.\cite{footnotephonons}
We denote 3D wavevectors by
upper case letters and their projection onto the 
$x$--$y$ plane by the corresponding lower case letter so that 
${\bf Q} = ({\bf q},Q_{z})$.
The electron creation (annihilation) operator in layer $i$ is  
$\hat{c}_{i}^\dagger({\bf k})$ ($\hat{c}_{i}({\bf k})$) with  
implicit spin indices,
the phonon creation (annihilation) operator for polarization $\lambda$ is 
$\hat{a}_{\lambda,{\bf Q}}^\dagger$ ($\hat{a}_{\lambda,{\bf Q}}$),
and the subband wavefunction of electrons in well $i$ is $\varphi_{i}(z)$.
With these definitions, the electron--phonon interaction contribution to
the Hamiltonian is given by 
\begin{eqnarray}\label{intham}
\hat{H}_{e-{\rm ph}} &=& V^{-1/2} 
\sum_{\lambda}\sum_{i=1,2} \sum_{{\bf q},Q_{z}}\
M_{\lambda}({\bf q},Q_{z})\, \hat{\cal A}_{\lambda,{\bf q},Q_{z}}\; \hat\rho_{i}(-{\bf
q})
F_{i}(Q_{z})
\end{eqnarray}
where ${V}$ 
is the normalization volume, 
\begin{eqnarray}
F_{i}(Q_{z}) &=& \int_{-\infty}^\infty\ dz\ |\varphi_{i}(z)|^2 e^{-iQ_{z}z}
\\
\hat{\cal A}_{\lambda,{\bf Q}} &=&
\hat{a}_{\lambda,{\bf Q}} +  \hat{a}^\dagger_{\lambda,-{\bf Q}}
\nonumber\\
\rho_{i}({\bf q}) &=& \sum_{\bf k} \hat{c}_{i}^\dagger({\bf k})
\hat{c}_{i}({\bf k}+{\bf q})
\end{eqnarray}
and  $M_{\lambda}({\bf Q})$ is the bulk electron-phonon coupling constant. 
At temperatures much lower than the Debye temperature one can 
neglect the Umklapp process in the electron-phonon interaction Hamiltonian.

\subsection{Kubo formula transconductivity}

The Kubo formula for linear response offers an expression for 
the trans{\em conductivity} tensor which is defined by
\begin{equation}
{\bf J}_{2}={\tensor\sigma}_{21} {\bf E}_{1}.
\end{equation}
In the absence of magnetic fields, ${\bf J}_{2}$ and ${\bf E}_{1}$ will
be anti-parallel, and ${\tensor\sigma}_{21}$ is a diagonal $2 \times 2$
tensor. 

The derivation sketched below for $\sigma_{21}$ is very similar to the one 
given previously in Ref.~\onlinecite{flen95a}, in which the reader 
can find further details.
The transconductivity is given by 
\begin{equation}
\sigma_{21}^{\alpha\gamma}({\bf k},\Omega)
=\frac{ie^2}{\hbar\Omega}\Pi_{21}^{\alpha\gamma,ret}
({\bf k},\Omega)
\end{equation}
where $\Pi_{21}^{\alpha\gamma,ret}({\bf k},\Omega)$ is the Fourier
transform of the retarded 
current-current correlation function, and $\alpha$ and $\gamma$ 
are Cartesian indices.
The retarded correlation function is evaluated by the
standard analytic continuation of the (bosonic Matsubara frequency)
Fourier components of its imaginary time counterpart.\cite{flen95a}
The imaginary time correlation function is calculated in 
perturbation theory
\begin{mathletters}
\begin{eqnarray}
\Pi_{21}^{\alpha\gamma}({\bf x}-{\bf x'},\tau-\tau')
&=& -\frac{\langle T_{\tau}\{ S(\beta) j_{2}^{\alpha}({\bf x},\tau)
j_{1}^{\gamma}({\bf x'},\tau')\}\rangle_{0}}{
\langle S(\beta)\rangle_{\rm 0}
},
\label{Pi=Sjj}\\
S(\beta)&=&T_{\tau}\left\{\exp\left[-\frac{1}{\hbar}
\int_{0}^{\hbar\beta}d\tau\ H_{\rm int}(\tau)\right]\right\},\label{smatrix}
\end{eqnarray}
\end{mathletters}
where $\langle \cdots \rangle_{\rm 0}$ denotes a non-interacting
system thermal average, and $T_{\tau}\{\cdots\}$ is the usual $\tau$-ordering
operator.  The $S$-matrix is expanded in powers of the interaction
Hamiltonian and the resulting non-interacting system 
correlation functions are evaluated with 
the aid of Wick's theorem.  As usual, the denominator in this expression
cancels the ``disconnected'' terms in the diagrammatic expansion.

If we include only electron-phonon interactions for the moment,
the lowest non-vanishing term appears at fourth order 
and gives the following contribution to the current-current 
correlation function:
\begin{eqnarray}
\Pi^{\alpha\beta}_{21}({\bf k}=0,i\Omega_{n})^{(4)}
&=&
-\frac{V^{-2} A^{-1}\hbar^{-4}}{4} 
\int_0^\beta d\tau \int_0^\beta d\tau_{1} \int_0^\beta d\tau_{2}
\int_0^\beta d\tau_3 \int_0^\beta d\tau_4\ \exp(i\Omega_{n}\tau)
\nonumber\\
&&
\sum_{{\bf Q}_{1},\lambda_{1}} \sum_{{\bf Q}_{2},\lambda_{2}} 
\sum_{{\bf Q}_{3},\lambda_{3}} \sum_{{\bf Q}_{4},\lambda_{4}}
M_{\lambda_{1}}({\bf Q}_{1})\, M_{\lambda_{2}}({\bf Q}_{2})\, 
M_{\lambda_3}({\bf Q}_3)\, M_{\lambda_4}({\bf Q}_4)
\nonumber\\
&&
\langle T_\tau\; \hat{j}^\alpha_{1}({\bf q}=0,\tau)\,
\hat\rho_{1}(-{\bf q_{1}},\tau_{1})\,
\hat\rho_{1}(-{\bf q_{2}},\tau_{2}) \rangle_{0}
\nonumber\\
&&
\langle T_\tau\; \hat{j}^\beta_{1}({\bf q}=0,0)\,
\hat\rho_{2}(-{\bf q_3},\tau_3)\,
\hat\rho_{2}(-{\bf q_4},\tau_4) \rangle_{0}
\nonumber\\
&&
F_{1}(Q_{z,1}) F_{1}(Q_{z,2}) F_{2}(Q_{z,3}) F_{2}(Q_{z,4})
\nonumber\\
&&
\langle T_\tau\;
\hat{\cal A}_{{\bf Q_{1}},\lambda_{1}}(\tau_{1})\,
\hat{\cal A}_{{\bf Q_{2}},\lambda_{2}}(\tau_{2})\,
\hat{\cal A}_{{\bf Q_{3}},\lambda_{3}}(\tau_{3})\,
\hat{\cal A}_{{\bf Q_{4}},\lambda_{4}}(\tau_{4})
\rangle_{\rm 0}
\label{Pi}
\end{eqnarray}
where $A$ is the 2D system area.   

The Wick's theorem factorization of the phonon operator product
expectation value leads to the product of two bare phonon Green's functions,
defined by 
\begin{equation}
D_{\lambda}^{(0)}({\bf Q},\tau-\tau')=
-\langle T_{\tau}\{\hat{\cal A}_{\lambda,{\bf Q}}(\tau)
\hat{\cal A}_{\lambda,-{\bf Q}}(\tau')\}
\rangle_{\rm 0}.
\end{equation}
{} It follows that, to leading order in electron-phonon interactions,
\begin{eqnarray}
\label{corr}
\Pi_{21}^{\alpha\gamma}({\bf k}=0,i\Omega)&=&\frac{-1}{2A\hbar^2}
\sum_{{\bf q}}\frac{1}{\hbar\beta}\sum_{i\omega}
\Delta_{2}^{\alpha}(-{\bf q},-{\bf q},-i\omega-i\Omega,-i\omega)
\Delta_{1}^{\gamma}({\bf q},{\bf q},i\omega+i\Omega,
i\omega)
\nonumber\\
&& \times
\int\frac{dQ_z} {2\pi\hbar} \sum_{\lambda}
F_{1}(Q_z)F_{2}(-Q_z)
|M_{\lambda}({\bf q},Q_z)|^2 
{D}_{\lambda}^{(0)}({\bf q},Q_z,i\omega) 
\nonumber\\
&&
\times\int \frac{dQ'_z}{2\pi\hbar}
\sum_{\lambda'}
F_{1}(Q'_z)F_{2}(-Q'_z)
|M_{\lambda'}({\bf q},Q'_z)|^2 
{D}_{\lambda'}^{(0)}({\bf q},Q'_z,i\Omega+i\omega) 
\end{eqnarray}
where
\begin{eqnarray}
{\bf\Delta}({\bf q},{\bf q};i\omega_{n},i\omega_{n}')
&\equiv& -{A}^{-1}
\int_0^\beta d\tau_{1} \int_0^\beta d\tau_{2}
\ \langle T_\tau
\hat{\bf j}({\bf q}=0,0) \hat\rho({\bf q},\tau_{1}) 
\hat\rho(-{\bf q},-\tau_{2})\rangle
\nonumber\\ 
&&\ \ \ \ \ \ \ \ \ \ \exp(i\omega_{n}\tau_{1})\;
\exp(i\omega_{n}'\tau_{2})
\label{define-Delta}
\end{eqnarray}
The Feynman diagram corresponding to this contribution to
the correlation function (\ref{corr}) is shown in Figure \ref{tria}. 

It will turn out to be important to account for disorder and 
anharmonicity in the lattice system.  We will do so using a 
phenomenological approach by introducing a phonon mean free path,
$\ell_{\rm ph}$. However, since the intrinsic bulk phonon mean free path may
exceed the dimensions of the sample, we should in principle take the
surface scattering explicitly into account. For simplicity we will
nevertheless use a single phenomenological mean free path and later, when
we discuss the long mean free path limit in more detail, take 
boundary effects into account.
The phonon Green's function is then\cite{mahan}
\begin{equation}
D_\lambda({\bf Q},i\omega_{n}) =
-\frac{2\omega_{\lambda,{\bf Q}}}
{(\omega_{n} + (c_{\lambda}/2\ell_{\rm ph})\,{\rm sgn}(\omega_{n}))^2
+ \omega^2_{\lambda,{\bf Q}}}
\end{equation}
where $\omega_{\lambda,{\bf Q}}=
c_{\lambda}\sqrt{q^2+Q_z^2}$, and $\ell_{\rm ph}$ is
the phonon mean free path. 

Note that the wavevector arguments in ${\bf \Delta}_i$ are 
2D, since ${\bf \Delta}_i$ is a property 
of the 2D electron  
systems. One can therefore sum over $\lambda$ and integrate over $Q_z$ to 
obtain a phonon-mediated effective interaction 
\begin{equation}
{\cal D}_{ij}({\bf q},i\omega_{n})
=\int\frac{dQ_z}{2\pi\hbar}\sum_{\lambda}
|M_{\lambda}({\bf Q})|^2 F_i(Q_z)F_j(-Q_z) D_{\lambda}
({\bf Q},i\omega_{n}).
\label{effint}
\end{equation}
This effective interaction is the 2D Fourier transform of the 
product of the phonon propagator between the layers and 
the electron-phonon interaction in each layer.
With this definition Eq.\ (\ref{corr}) becomes
\begin{eqnarray}
\Pi_{21}^{\alpha\gamma}({\bf k}=0,i\Omega)&=&\frac{-1}{2A\hbar^2}
\sum_{{\bf q}}\frac{1}{\hbar\beta}\sum_{i\omega}
\Delta_{2}^{\alpha}(-{\bf q},-{\bf q};-i\omega-i\Omega,-i\omega)
\Delta_{1}^{\gamma}({\bf q},{\bf q};i\omega+i\Omega,
i\omega)
\nonumber\\
&& {\cal D}_{21}({\bf q},i\omega_{n})
{\cal D}_{21}({\bf q},i\Omega_{n} + i\omega_{n})
\end{eqnarray}
This expression for $\Pi_{21}$ is the same as in 
Ref.~\onlinecite{flen95a}, except that the interlayer Coulomb
interaction is replaced by the phonon-mediated effective interaction.
{\null} From this point on, the formal steps are identical to the
Coulomb case.  Performing the summation over $i\omega$, continuing to 
real frequencies and taking the $\Omega\rightarrow 0$ limit,
we obtain\cite{flen95a}
\begin{eqnarray}\label{tcon}
\sigma_{21}^{\alpha\gamma}&=&\frac{e^2}{2\hbar^3 A}\sum_{{\bf q}}
\int_{-\infty}^{\infty}\frac{d\omega}{2\pi}\ 
|{\cal D}_{21}({\bf q},\omega+i\delta)|^2
\left[-\frac{\partial n_{B}(\omega)}{\partial \omega}\right]
\nonumber\\
&&\times
\Delta_{2}^{\gamma}(-{\bf q},-{\bf q},-\omega-i\delta,-\omega+i\delta)
\Delta_{1}^{\alpha}({\bf q},{\bf q},\omega+i\delta,\omega-i\delta).
\end{eqnarray}

\subsection{From Transconductivity to Transresistivity} 

For further progress it is necessary to make some assumptions
about the electronic systems.  We will assume that the 
2D electron layers are good metals with large $k_{F} \ell$
where $\ell$ is the electronic mean free path.
It can then be shown\cite{kame95,flen95a,flen95b}
that whenever the transport scattering time, $\tau_{tr}$, is independent of 
energy, the function $\Delta$ is related to the electron polarization
function $\chi({\bf q},\omega)$ by 
\begin{equation}\label{rel}
\Delta_{i}^{\alpha}=\frac{2\tau_{tr,i}}{m^*}q^{\alpha}
{\rm Im}\chi_{i}({\bf q},\omega).
\end{equation}
Here $m^*$ is the electron effective mass. 
The relation (\ref{rel}) is a property of the 2D 
electron layers only, and is not dependent on the phonon degrees of freedom.
Assuming further that $|\sigma_{12}| \ll \sigma_{ii} $ the 
transresistivity can be approximated as follows:  
\begin{equation}
\rho_{21}=\frac{-\sigma_{21}}{\sigma_{11}\sigma_{22}-\sigma_{12}
\sigma_{21}}\approx\frac{-\sigma_{21}}{\sigma_{11}\sigma_{22}}.
\end{equation}
Using the Drude expression ($\sigma_{ii}=e^2n_{i}\tau_{tr_{i}}/m^*$ )
for the intralayer conductivities, which is  
valid under the above assumptions, the transresistivity 
due to the electron--phonon interaction is given by the following 
explicit expression,
\begin{equation}\label{rho}
\rho_{21} = \frac{-\hbar^2}{4e^2 n_{1} n_{2} k_{B}T}\frac{1}{A}\sum_{{\bf q}} 
q^2 \int_{-\infty}^\infty\frac{d\omega}{2\pi}
\left|{\cal D}_{21}({\bf q},\omega) \right|^2
\frac{{\rm Im}\chi_{1}({\bf q},\omega)
{\rm Im}\chi_{2}({\bf q},\omega)}
{\sinh^2(\hbar\omega/2k_{B}T)}.
\end{equation}

The same expression for the phonon-exchange contribution to the 
drag can also be derived using semiclassical Boltzmann transport
theory and a collision term with transition matrix elements 
calculated by summing over virtual and real intermediate states with 
absorbed and emitted phonons.  Ambiguities can arise in
that approach, however, from the portion of phase space 
where the relevant energy denominators approach zero.  As we 
explain below this part of the phase space {\em is} important in 
determining the drag resistivity.  Our Kubo function derivation
allows finite phonon mean-free-paths, which remove any
spurious singularities, to be incorporated into the 
calculation in a consistent and unambiguous manner.  

Note that the explicit dependence of the 
transconductivity on the transport lifetime is absent 
in the transresistivity.  (An implicit dependence
remains through the dependence of the polarization function 
on disorder.)  This aspect of the final expression has 
been emphasized in recent work of Swierkowski, Szyma\'{n}ski
and Gortel,\cite{ssgrecent}
is not accidental, and emerges naturally
in a force-balance approximation where the drag force 
simply cancels the rate of momentum transfer per particle from
the current carrying layer to the open layer.  

\subsection{Coulomb interaction and screening}

In reality, electron--electron and electron-phonon interactions 
are simultaneously present and both should be included 
in a drag calculation.  To leading order, Coulomb interactions  
can be incorporated by simply adding\cite{squares} the interlayer Coulomb
interaction to the phonon-mediated effective interaction in
Eq.~(\ref{rho}).   One class of higher order terms in which intralayer
Coulomb interactions appear is captured 
by replacing the electronic polarization functions 
which appear in Eq.~(\ref{rho}) by their interacting system
counterparts.  However,
the most essential higher order terms are those which account 
for the screening of both phonon-mediated and Coulombic interactions.
In the RPA, 
the total interlayer screened interaction is given by 
\begin{equation}\label{w-12}
W_{21}^{\rm total}(q,\omega) = \frac{{\cal D}_{21}(q,\omega)+U_{21}(q)}
{\epsilon(q,\omega)},
\end{equation}
where $U_{ij}(q)$ is the unscreened Coulomb interaction and
\begin{equation}\label{eps}
\epsilon(q,\omega) =  [1 - ({\cal D}_{11}+U_{11})\chi_{1}]
[1 - ({\cal D}_{22}+U_{22})\chi_{2}]
- ({\cal D}_{21}+U_{21})^2 \chi_{1} \chi_{2}.
\label{epsilon}
\end{equation}
(See for example Ref.~\onlinecite{zhan93}. The  
form of the phonon-mediated interlayer interaction in this
reference is incorrect, however.)  $\epsilon(q,\omega)$ is the 
effective dielectric function for interlayer interactions in the 
RPA.  Notice that when the contribution of interlayer interactions 
to the screening can be neglected, $\epsilon$ is simply the 
product of the dielectric functions for the two layers, corresponding 
to independently screened electron-phonon interactions in each layer.

Coulomb and phonon-mediated interactions can be simultaneously 
included in the transresistivity simply by replacing
${\cal D}$ by $W^{\rm total}$ in Eq.~(\ref{rho}).
Note that the transresistivity is not strictly the sum of 
purely ``Coulomb'' and ``phonon'' contributions, since  
there are interference terms proportional to $U \times {\cal D}$ in 
the $|W_{21}^{{\rm total}}|^2$.  However, the Coulomb contribution
is large only for $q \lesssim 0.5 k_{F}$, 
($U(q) = 2 \pi e^2 \exp (-q d) /q $ when the finite thickness of
the electron layers is neglected, 
whereas contributions from the ${\cal D}$ term come predominantly from 
$q \approx 2 k_{F}$.  Hence, the interference
terms are usually negligible, and in practice we will treat 
the Coulomb and phonon contributions as if they were incoherent. 
In what follows, we will concentrate on the ``phonon
contribution,'' which will be calculated from Eq.~(\ref{rho}), with  
only the ${\cal D}$ term in the numerator, 
\begin{equation}
W_{21}(q,\omega) = \frac{{\cal D}_{21}}{\epsilon}.
\end{equation}
Note that it is important to retain the coupling between the 
Coulomb and phonon terms in $\epsilon$, since this can influence the 
$q \approx 2 k_{F}$ contribution to the transresistivity.

\section{The phonon mediated interaction in GaAs/Al$_{x}$Ga$_{1-x}$As systems}
\label{inter}

As mentioned earlier, the acoustic electron--phonon interaction 
is weak in GaAs and ${\rm Al}_{x}{\rm Ga}_{1-x}{\rm As}$. 
In this section we discuss quantitatively what ``weak'' means, and 
explain how phonons can give an important contribution despite 
this weakness.  In the following section we discuss how and 
when screening affects the phonon mediated interaction. 

\subsection{Electron-phonon coupling in GaAs/AlGaAs}  

In GaAs/AlGaAs systems, electrons couple to acoustic
phonons via deformation potential and piezoelectric couplings. 
Since we are concerned with low energy excitations, only acoustic phonons 
in the long wavelength limit have to be considered. 
In this limit, the squares of the {\em e--ph} coupling 
strengths for longitudinal and transverse phonons are,\cite{pjprice}
respectively,
\begin{eqnarray}
|M_{l}({\bf Q})|^2 &=&
\frac{\hbar Q}{2\varrho c_{l}}
\left[D^2+ \frac{(eh_{14})^2}{Q^2} A_{l}({\bf Q})\right]
\label{m-1}\\
|M_{t}({\bf Q})|^2 &=& \frac{\hbar(eh_{14})^2}{2\varrho c_{t} Q}A_{t}({\bf Q})
\label{m-t}
\end{eqnarray}
where $\varrho$ is the mass density of the crystal, $D$ is 
the deformation potential, $eh_{14}$ is the piezoelectric constant, 
the $c_{\lambda}$ are sound velocities for longitudinal and
transverse phonons, and the $A_{\lambda}$ are 
the anisotropy factors,\cite{lyo88}
\begin{mathletters}
\begin{eqnarray}
A_{l}({\bf Q}) &=& \frac{9q^4Q_z^2}{2Q^6},
\label{a-l}\\
A_{t}({\bf Q}) &=& \frac{8q^2Q_z^4+q^6}{4Q^6}.
\label{a-t}
\end{eqnarray}
\end{mathletters}

\subsection{Approximate analytic form of ${\cal D}({\bf q},\omega)$}
\label{approx-anal}

While it is possible to obtain exact expressions for ${\cal D}$ 
from Eq.\ (\ref{effint}) including the full anisotropy functions
(Eqs. (\ref{a-l}) and (\ref{a-t})) and
the form factors for infinite square wells, these are extremely 
complicated.  Therefore, we shall make some well-controlled approximations
explained below which do not significantly affect the final results for 
the computed transresistivity. 

As we shall see shortly, the phonon-mediated effective interactions are 
important only when $\omega$ is close to $c_\lambda q$.  
For these $\omega$'s the integral over $Q_z$ in Eq.\ (\ref{effint}) 
is dominated by contributions from near $Q_z =0$.  
We therefore remove all $Q_z$ dependent factors, except for the small
energy denominator, from the integral which defines ${\cal D}$.
For example, one can set $Q_z=0$ in the anisotropy factors given in 
Eq.\ (\ref{a-l}) and (\ref{a-t}), yielding 
\begin{equation}
A_{l} \approx 0,\ \ \ \ A_{t} \approx \frac{1}{4}.
\end{equation}
Then the phonon-mediated effective interaction 
for two equivalent infinite square wells with width $L$ and
center-to-center separation of $d$ is 
\begin{eqnarray}
\nu_0 \,{\cal D}_{ij}(q,\omega) &\approx& - \frac{3 C_{DP}}{k_{F} L}
\delta_{ij}
 - C_{DP}\;\frac{q\zeta_l^2}{k_{F}\sqrt{1 - \zeta_l^2}}\;
B_{ij}\left( q d \sqrt{1 - \zeta_l^2},
q L \sqrt{1-\zeta_l^2}\right)
\nonumber\\
&& \ - C_{PE} \frac{k_{F}}{q\sqrt{1 - \zeta_{t}^2}}\;
B_{ij}\left(q d\sqrt{1 - \zeta_{t}^2}, q L \sqrt{1-\zeta_{t}^2}\right),
\label{gD_{i}j}
\end{eqnarray}
Here, $\nu_0 =m^*/\pi\hbar^2$, the 2-dimensional electron 
gas density of states (so that $\nu_0{\cal D}$ is dimensionless), 
\begin{eqnarray}
\zeta_{\lambda} &=& \frac{\omega}{c_{\lambda} q} + \frac{i}{2 q \ell_{\rm ph}}
\label{defzeta}
\\
C_{PE} &=& \frac{(eh_{14})^2 m^*}{8\pi\hbar^2 c_{t}^2\varrho
k_{F}}
\\
C_{DP} &=& 
\frac{D^2 m^* k_{F}}{2 \pi \hbar^2 c_{l}^2 \varrho}
\label{cdp}
\end{eqnarray}
and 
\begin{equation}
B_{ij}(x,y)=\left\{
\begin{array}{ll}
\frac{\pi^2}{y^2+\pi^2}\left(\frac{3y}{2\pi^2}+\frac{1}{y}
+\frac{1}{2y^2}\frac{\pi^2}{y^2+\pi^2}
(e^{-2y}-1)\right)\ &  ,i=j; \\
\exp(-x)\ \left(\frac{\pi^2}{y^2+\pi^2}\right)^2
\frac{\sinh^2(y)}{y^2} & ,i \neq j.
\end{array}
\right.
\label{B_ij}
\end{equation}
In the expressions above, the square root with a positive 
real part should be taken.

Inserting numerical values for GaAs ($m^* = 0.067 m_e$, $c_l = 5.14 \times
10^5\,{\rm cm/s}$, $c_t = 3.04 \times 10^5\,{\rm cm/s}$,
$\varrho = 5.3\,{\rm g/cm^3}$, $eh_{14} = 1.2 \times 10^7\,{\rm eV/cm}$ and
$D = -13.0\,{\rm eV}$), gives the following dimensionless coupling constants:
\begin{eqnarray}
C_{PE} &\approx& 
1.64 \times 10^{-3} \times \frac{10^6\,{\rm cm}^{-1}}{k_{F}}\\
C_{DP} &\approx& 2.7 \times 10^{-3} \times 
\frac{k_{F}}{10^6\,{\rm cm}^{-1}} 
\end{eqnarray}

For the same subband wavefunctions the bare interlayer and intralayer 
Coulomb interactions are given by 
\begin{equation}
\nu_{0} \, U_{ij}(q) = \frac{q_{\rm TF}}{q}\; B_{ij}(qd,qL),
\label{coulomb}
\end{equation}
where $q_{\rm TF}=2\pi e^2\nu_0$ is the Thomas Fermi wavevector. (We have 
absorbed the bulk dielectric constant of the semiconductor in the 
electron charge).

\subsection{Strength of ${\cal D}_{21}$}

For the RPA screened Coulomb interaction ($U_{\rm RPA}$) 
in a single layer system, 
the magnitude of $\nu_0 U_{\rm RPA}$ approaches $1$ at long wavelengths.
Although the corresponding value for double-layer systems is 
smaller,\cite{gram-coul} it is useful to compare $\nu_0 {\cal D}_{21}$ 
to this value. 
For our present illustrative purpose we concentrate on 
the deformation potential term ({\em i.e.} the $D^2$ term in 
Eq.\ (\ref{m-1})), which turns out to dominate except for very
low density electronic systems.  Then, in the limit of 
vanishing quantum well widths and $\ell_{\rm ph} \to \infty$, 
we obtain
\begin{eqnarray}
\nu_0 {\cal D}_{21}(q,\omega+i\eta) &\approx& 
- C_{DP}\;\frac{\omega^2}{q k_{F} c_l^2\sqrt{1 - \omega^2/(q c_l)^2}}\;
\exp\left(-d\sqrt{q^2 - \omega^2 c_l^{-2}}\right)
\label{D-approx}
\end{eqnarray}
Note that the magnitude of the effective interaction
{\sl diverges} as $q \rightarrow \omega /c_{l}$ from above or below.  
(We point out that the transresistivity is 
strongly dependent on the width $L$, and therefore one should not 
take the $L\rightarrow 0$ limit when computing ${\cal D}$.
We use the full form given by Eq.\ (\ref{gD_{i}j})  
in subsequent numerical calculations.)

The small prefactor $C_{DP}$ guarantees that the 
phonon-mediated effective interaction is small compared 
to the Coulomb interaction except near $\omega \approx q c_{\lambda}$.
The large value of the interaction in this region of phase space
reflects the large phase space for intermediate states with 
small or vanishing energy denominators when $\omega \approx q
c_{\lambda}$.  (Note that the phonon energy varies slowly with
$Q_z$ for $Q_z$ near zero.)  

The importance for drag of the sharp peak in ${\cal D}_{21}({\bf q},\omega)$
is enhanced by the fact that it appears squared in Eq.\ (\ref{rho}). 
As a result, phonons do play an important role in frictional drag, 
even though the typical value of $\nu_0 {\cal D}_{21}({\bf q},\omega)$
is small.  In fact, if we ignore the effects of screening and let 
$\ell_{\rm ph}\rightarrow\infty$, we get an infinite transresistivity. 
This is easily seen from Eq.~(\ref{D-approx}).
The absolute value ${\cal D}_{21}$ diverges like 
$(c_{\lambda}q-\omega)^{-1/2}$ as $\omega$ 
approaches $c_{\lambda}q$ from both above and below. 
Inserting this bare form of the interaction 
into the Eq.\ (\ref{rho}) gives a $|c_{\lambda} q - \omega|^{-1}$
divergence in the energy transfer integral 
for every $q$, yielding an {\em infinite} transresistivity. 
This point does not seem to have been emphasized in the 
existing theoretical literature on this subject.  
As we show in the next two subsections, including either 
a finite mean free path or screening the interaction dynamically 
removes this spurious divergence. 

In Fig.~\ref{wplot} we plot
$\nu_0|{\cal D}_{21}(2k_F,\omega)/\epsilon(2k_F,\omega)|$ near
the longitudinal resonance for different values of $\ell_{\rm ph}$.
It is useful to compare $\nu_0|{\cal D}_{21}/\epsilon|$ to the
screened interlayer Coulomb interaction $\nu_0U_{21,{\rm RPA}}
\propto \exp(-q d)/(1 + q_{\rm TF}/q)$. 
In contrast to the phonon-mediated interaction near the resonance, 
$\nu_0U_{21,{\rm RPA}}$ decreases rapidly below unity as a function of 
well separation. Fig.~\ref{wplot}, partly
explains how the ``weak'' phonon-mediated interaction can compete
with the Coulomb interaction as a mechanism for drag.

\subsection{Effect of Screening}
\label{screening}

{}From Eq.\ (\ref{gD_{i}j}), one sees that the presence of a
finite $\ell_{\rm ph}$ cuts off the divergence and leads to
a finite transresistivity.  Since the divergence in the
integrand is of the form $1/|\omega - c_{\lambda} q|$, it follows 
that $\rho_{21}$ would be proportional to $\ln(\ell_{\rm ph})$
if screening were not important.  When screening is included
the drag resistivity does not diverge. 

As shown in Eqs. (\ref{epsilon}) and (\ref{w-12}),
screening is accounted for by dividing the bare interlayer interaction 
${\cal D}_{21}(q,\omega)$ by a dielectric function 
$\epsilon(q,\omega)$.  
Whenever ${\cal D}_{21}(q,\omega)$ diverges at $\omega
= c_{\lambda} q$, so does $\epsilon(q,\omega)$.  The screened
interaction is therefore nondivergent, even when 
$\ell_{\rm ph}\rightarrow\infty$.

As we explain below, screening becomes important for the 
phonon-exchange drag only if the 
phonon mean free path exceeds a critical value 
$\ell_{\rm ph,crit}$. Since $\ell_{\rm ph,crit}$ is close to realistic 
values, we investigate the two regimes separately in the following two sections.

\section{Long mean free path limit}\label{couplemode}

In this section we focus on the large phonon mean free path limit, where
a coupled electron-phonon mode turns out to be of utmost importance.
First we discuss the ideal case of infinite $\ell_{\rm ph}$, where the
coupled mode is broadened by the coupling to the electronic system only.
In this case we find an analytic form for the coupled mode contribution.
Secondly, in Section~\ref{longbutfinite}, we discuss how the collective
mode contribution is modified by a finite mean free path and the
conditions for its experimental observation. 

\subsection{Infinite phonon mean free path limit}\label{infinite}

The approximate analytic results discussed 
below apply only for layer separations smaller than
a large but finite maximum value which we specify below; for still larger
layer separations the analytic analysis is less revealing and 
we have relied more strongly on numerical studies.  

We show below that 
for $\ell_{\rm ph} \to \infty$, the real part of $\epsilon(q,\omega)$ 
vanishes and the imaginary part is small enough to 
yield a sharp collective mode for $\omega$ just below 
$ q c_{l} $ which contributes strongly to the 
frictional drag between the two layers.  
To make the following discussion as transparent as possible 
we limit our attention 
to the case of identical electron layers 
so that ${\cal D}_{22} = {\cal D}_{11}$ and $U_{11} = U_{22}$.
We locate the collective mode frequency $\omega_{0}(q)$ 
by solving the equation Re$[\epsilon(q,\omega_{0})] = 0$.
We will be interested in 
momentum transfers near $2 k_{F}$ which give the 
main contribution to the drag and systems with $k_{F} d \gg 1$ 
so that we can neglect $ U_{21}(q) \propto \exp ( - 2 k_{F} d )$.
The small layer separation approximation mentioned above consists
of setting ${\cal D}_{11} \approx {\cal D}_{12}\equiv {\cal D}$ in 
Eq. ~(\ref{eps}) which is justified for 
\begin{equation}
d \ll \frac{1 + q_{\rm TF}/2 k_{F}}{16 C_{\rm DP} k_F}\equiv d_B
\label{d_{l}imit}
\end{equation}
as we show later. With these assumptions
\begin{eqnarray}
W_{21}(q,\omega) &\approx&
\frac{{\cal D}_{21}(q,\omega)}
{(1 - U_{11}\chi)(1 - U_{11}\chi -2 {\cal D}_{} \chi)} 
\label{w-21approx}
\end{eqnarray}
The collective mode in which we are interested occurs 
near  $ \omega = c_{l} q \ll v_{F} q$ and hence is in the 
low frequency regime for $\chi ( q, \omega)$.  Here
$v_{F}$ is the Fermi velocity of the electronic system.  We therefore
approximate ${\rm Re}\chi ( q, \omega) \approx \chi (q,\omega =0) 
= - \nu_{0}$.  We write the imaginary part of the polarization 
function in the form ${\rm Im} \chi (q , \omega ) 
= -\nu_{0} \tilde\delta$.   In the low frequency limit of the RPA, 
$ \tilde\delta (q,c_{l} q) \approx c_{l} / (v_{F} 
\sqrt { 1 - q^2 / 4 k_{F}^2})$, except near $2k_F$ where   
it is approximately $\sqrt{c_l/2v_F}$.
$\tilde\delta$ is dimensionless
and has a wavevector and frequency dependence which will be left
implicit in the following discussion except where emphasis is
important.  However, it is important for the following discussion that 
$\tilde\delta(q,c_lq)$ is generally small compared to one. 

Because we are interested in large wavevectors near $q = 2 k_{F} $ 
we include only longitudinal deformation potential coupling. 
The calculation including transverse piezoelectric mode follows
{\em mutatis mutandis}.  From the previous section, 
we have, for $\omega$ close to but smaller than $ c_{l} q $, \, 
\begin{equation}
\nu_{0} W_{21}(q, \omega) = \frac{ - C_{DP} }
{\epsilon_{0} ( \frac{k_{F}}{q}\epsilon_{0} \sqrt{ 1 - \omega^2 / c_{l}^2 q^2 } - 
2 C_{DP}) - i 2 C_{DP} \tilde\delta }.
\label{largelphapprox} 
\end{equation} 
In this equation $\epsilon_{0} = 1 + q_{\rm TF} /q $
is the RPA static dielectric function for an isolated layer.
The collective mode occurs where the real part of denominator
of Eq.~(\ref{largelphapprox}) vanishes at
\begin{equation}
\omega_{0} = c_{l} q \sqrt{1 - \left(\frac{2qC_{DP}}{k_{F}\epsilon_{0}}\right)^2}
\label{collectivemode}
\end{equation} 
This collective mode results from coupling of the electron layers
to phonons with $\omega$ close to $c_{l} q$. 
Expanding the denominator of 
Eq.~(\ref{largelphapprox}) around the pole we find that for
$\omega$ close to $\omega_{0}$  
\begin{equation}
\big| \nu_{0} W_{21}(q,\omega) \big |^2 = 
\frac{c_{l} q^3 C_{DP}^2} { \epsilon_{0}^3k_{F}^2 \tilde\delta} 
\frac{ \Gamma } { (\omega - \omega_{0})^2 + \Gamma^2 }
\label{lorentzian} 
\end{equation} 
where the width of the collective mode resonance is given by 
\begin{equation}
\Gamma  = \frac{ 4 \tilde\delta c_{l} q^3 C_{DP}^2  } 
{ \epsilon_{0}^3 k_{F}^2}.
\label{gamma}
\end{equation} 
Note that the width of the resonance is small compared to the 
shift of the resonance from $c_{l} q$ and that the resonance line 
shape is approximately Lorentzian only if $\tilde\delta \ll 1$.  
Where this condition is not satisfied Eq.~(\ref{lorentzian}) 
will not be accurate.

Numerical calculations discussed later demonstrate that 
for $\ell_{\rm ph} \to \infty$ the drag is dominated by 
coupling associated with this collective mode resonance.  
Since the effective interaction has a more
rapid frequency variation than other quantities in the expression for
the transresistivity (Eq.~(\ref{rho})) 
we may approximate the screened interaction near the resonance 
by a $\delta$ function:
\begin{equation}
\big| \nu_{0} W_{21}(q,\omega) \big|^2 = 
\frac{ c_{l} q^3 C_{DP}^2  } { k_{F}^2 \epsilon_{0}^3 \tilde\delta}\pi
\delta ( \omega - \omega_{0}) 
\label{deltafunction} 
\end{equation} 
This approximation allows the frequency integral in 
Eq.~(\ref{rho}) to be performed and the contribution from 
the coupled mode to the transresistivity may be expressed as 
\begin{equation}\label{long-mfp-approx}
\rho_{21,\lambda} \approx \left(\frac{h}{e^2}\right) 
\frac{\pi K_\lambda}{n^2 k_BT} \int_0^\infty dq
\left(\frac{q^\alpha}{\epsilon_0^3}\right)
\frac{\nu_0 {\rm Im}\chi(q,c_\lambda q)}
{\sinh^2(\hbar c_{\lambda}q/2k_{B}T)}.
\end{equation}
where $\alpha = 6$ for the longitudinal phonons and
$\alpha = 2$ for the transverse phonons, and where
\begin{mathletters}
\begin{eqnarray}\label{longcst}
K_l&=&\frac{\hbar D^4}{(4\pi)^3 c_l^3 \varrho^2},\\
K_t&=&\frac{\hbar (eh_{14})^4}{2^{10}\pi^3 c_t^3 \varrho^2},\\
\end{eqnarray}
\end{mathletters}
The $q^6/\epsilon_0^3$ term in the integrand above implies that for 
$T \gtrsim T_{\rm BG}=2\hbar c_l k_F/k_B$,
the main contribution to the integral comes from the $q = 2k_F$ 
region; {\it i.e.} large-angle scattering dominates the phonon
mediated drag. 

Comparing with Eq.~(\ref{coul}) we see that, at least for 
$\ell_{\rm ph} \to \infty$, the phonon mediated drag can be 
comparable to or stronger than Coulomb drag, in spite of the weak 
electron phonon interactions.  
Crudely the condition which needs to be satisfied is that 
$T_{\rm BG}$ is low enough or the layer separation is large enough that 
$ C_{DP}^2 \sqrt{c_l/v_F}
\gtrsim  (k_{B} T_{\rm BG} / \epsilon_{F})^2 (q_{\rm TF} d)^{-2} 
( k_{F} d )^{-2}$.
For typical layer densities in GaAs this condition is satisfied for 
layer separations larger than a few tens of nanometers, 
consistent with experimental observations. 
When phonon mediated drag is dominant, $\rho_{21}$ will be proportional
to temperature for $T\gg T_{\rm BG}$. Below $T_{\rm BG}$ there will be 
a crossover to a regime where the piezoelectric contribution dominates
and the temperature dependence goes approximately 
as $T^5$. This can be seen from Eq.~(\ref{long-mfp-approx}) by defining 
a new integration variable proportional to $q/T$ and taking the 
$q$-dependence of $\epsilon_0$ into account. 
We shall discuss 
this further in Section~\ref{tempdep}. At extremely low temperatures,
the assumption that the drag rate is determined 
by the resonance $\omega\simeq c_lq$ will break down. $\rho_{21}$ is 
then dominated by the $\omega=0$ limit of ${\cal D}_{21}$ and will 
revert to the familiar $T^2$ law for carrier-carrier scattering in a 
Fermi liquid. 

Recall that we used ${\cal D}_{21} \approx {\cal D}_{11}$ to obtain the
above results.  From Eq.\ (\ref{gD_{i}j}), 
\begin{equation}
\frac{{\cal D}_{21}^2(q,\omega_{0,l}(q))}
{{\cal D}_{11}^2(q,\omega_{0,l}(q))}
\approx \exp(- 2qd \sqrt{2 (1 - \omega_{0} / c_{l} q) }).
\label{compare}
\end{equation}
Hence, for the approximation ${\cal D}_{11}^2 - 
{\cal D}_{21}^2 \approx 0$ to be valid, one must have 
$2q d \sqrt{2 (1 - \omega_{0} / c_{l} q) }   \ll 1$. 
Together with Eq.\ (\ref{collectivemode}), this gives the
condition Eq.\ (\ref{d_{l}imit}).  
For $\ell_{\rm ph} \to \infty$, the preceding approximate
calculation implies that there is no layer separation dependence 
until this length, which for GaAs and typical densities 
corresponds to $d \sim 5000$ \AA, is reached.   
Numerical results show, however, that there is a weak 
distance dependence on the transresistivity in this large $\ell$
regime.  This stems from the presence of a relatively long 
non-Lorenzian tail for $\omega < c_l q$ which contributes significantly 
to the integral, as we will discuss below.  

At larger layer separations, the interlayer phonon propagator 
${\cal D}_{21}$ 
is reduced at the collective mode frequency and the two
electron layers interact with the phonon system more 
independently.  We must then use a more refined expression for
the effective interaction: 
\begin{eqnarray}
W_{21}(q,\omega) &\approx&
\frac{{\cal D}_{21}(q,\omega)}
{(1 - [U_{11}+ {\cal D}_{11}]\chi)^2  - {\cal D}_{21}^2 \chi^2}
\end{eqnarray}
A competition occurs between a decline 
in the coupling due to suppression of the interlayer
propagator ${\cal D}_{21}$ and resonant enhancement of 
the electron phonon interaction near each electron layer.

\subsection{Coupled electron-phonon mode with finite mean free path}
\label{longbutfinite}

The above analysis is based on phonons with an infinite mean free path,
apart from the finite lifetimes due to interactions with the 
electronic layers.
Any real system will have imperfections which will make 
the phonon life time finite even when the electron layers are not present.
Even when the lattice is perfect and free of 
isotopic impurities, anharmonicity and boundary scattering 
will cause phonon modes to decay.  In the following 
we represent all these effects in the simplest possible
way by assigning a common phenomenological mean free path $\ell_{\rm ph}$
to all modes.

The occurrence of coupled phonon-plasmon collective modes, signaled by 
by a zero in the ${\rm Re}[\epsilon (q,\omega)]$, requires a 
cancellation between phonon-mediated and Coulomb interaction 
contributions.
In the present section we will consider the limit in which 
the phonon-mediated contributions to $\epsilon (q,\omega)$ can
be neglected, {\em i.e.} the limit in which 
$|\nu_{0} {\cal D}| \ll 1$ for almost all energy and 
momentum transfers and for both intra-layer and inter-layer
propagation.  We will see that this condition is satisfied 
except at long phonon mean free paths.

For $\omega$ close to $ q c_{l}$ and finite $\ell_{\rm ph}$ 
(but large compared to $q^{-1}$), 
\begin{equation}
\left|\nu_{0}{\cal D}_{21}\right| \approx 
\left(q C_{DP}/{k_F}\right)
|\sqrt{1-\zeta_{l}^2}|^{-1}.
\end{equation}
where $\zeta_l$ is defined in Eq.~(\ref{defzeta}), and
$1 - \zeta_l^2 \approx 1 - (\omega/q c_l)^2 + i/q\ell_{\rm ph}$.
There is a critical value $\ell_{\rm ph,crit}$ such 
that $1-(\omega_{0}/c_{l}q)^2\lesssim 1/q\ell_{\rm ph,crit}$, 
in which case $|{\rm  Re}[{\cal D}_{ij}]|$ 
never becomes large enough for ${\rm Re}[\epsilon] = 0$ to have 
a solution and the collective mode ceases to exist. 
This critical value is given by 
\begin{equation}
\ell_{\rm ph,crit} = \frac{\epsilon_{0}^2k_{F}^2}{4C_{\rm DP}^2 q^3}
=\frac{(1+q_{\rm TF}/2k_{F})^2}{32 C_{\rm DP}^2 k_{F}}, 
\end{equation}
where, since large momentum transfers dominate the drag,
we have set $q=2k_{F}$.  $\ell_{\rm ph,crit}$ 
is even longer at smaller $q$'s and it seems improbable 
that the collective mode discussed in the previous section 
would ever be evident in inelastic light scattering studies
of long-wavelength electronic excitations. 
In GaAs, and for $k_{F} \approx 10^{6}\,{\rm cm}^{-1}$, the critical 
mean free path is approximately $\ell_{\rm ph,crit}\approx 0.2$ mm.
The actual mean free path of course depends on the sample in question. 

It is often the case that scattering of phonons off the boundaries 
of the sample can be accounted for by taking the mean free path 
in the absence of bulk scatterers to be equal to the sample size.
We argue below that for the present problem, the sample size in the 
$z$-direction is usually irrelevant and that the maximum mean free 
path is given by the typically larger lateral dimension of the sample.

The phonons which contribute to the coupled mode are those which 
are confined around the two-dimensional electron gases.  
The extent of the confinement is given by range ${\cal D}_{21}$ 
in the $z$-direction ({\it i.e.}, perpendicular
to the 2DEGs). To study the $z$-direction range of the phonon field  
participating in the collective mode, we look more closely at which
$z$-region actually contributes to the coupling.  
We write the effective phonon
mediated interaction given in Eq.\ (\ref{effint}) as 
\begin{equation}
D_{21}(q,\omega+i\delta) = \int dzdz'\ 
|\varphi_2(z)|^2|\varphi_1(z')|^2 K(q,\omega+i\delta,z-z'),
\label{D12zz}
\end{equation}
where $K(q,\omega+i\delta,z-z')$ is the Fourier transform of the 
phonon Green's function
and the electron-phonon coupling matrix element
with respect to $Q_z$,
\begin{equation}
K(q,\omega+i\delta,z) = \sum_\lambda
\int \frac{dQ_z}{2\pi\hbar} e^{iQ_zz}
D_\lambda({\bf Q},\omega+i\delta) |M_\lambda({\bf Q})|^2.
\end{equation}
For a given $q$ and $\omega$, the spatial extent in the $z$-direction 
of $K(q,\omega+i\delta,z)$ gives the range of the phonon field.
Since the collective mode frequencies are very close to $\omega=c_lq$, 
the relevant phonon wavevectors in the $z$-direction 
are small, and the same approximations used previously can also be 
applied here, yielding an approximate form for $K$
given by Eq.\ (\ref{D-approx}) with $d$ replaced by $|z-z'|$; 
{\em i.e.}, $K(q,\omega+\delta,z-z')\sim 
\exp(-|z-z'|\sqrt{q^2-\omega^2/c_l^2})$. 
The spatial extent of the phonon field participating in 
the collective mode can be found by substituting the frequency for
mode, $\omega_0$, into this exponential form, 
giving $K\sim \exp(-|z-z'|/d_B)$, where 
$d_B$ was defined previously in Eq.~(\ref{d_{l}imit}). 
For GaAs at $n=1.5\times 10^{11}\,{\rm cm}^{-2}\ $ $d_B$ is approximately
5000 \AA.  
Any boundaries or imperfections beyond this range in the $z$-direction 
have negligible effect.

It is plausible that some existing  
experimental results are for samples with 
$\ell_{\rm ph}$ in the lateral direction 
comparable to this critical mean free path
and hence partially reflect collective excitations of the 
electron-phonon system. However, more experiments are needed before
definite conclusions can be drawn.

\section{Short phonon mean free path limit}\label{phdamped}

In the present section, we will consider the limit in which phonon mediated
contributions to $\epsilon(q,\omega)$ can be neglected, {\it i.e.}, 
$\ell_{\rm ph} \ll \ell_{\mathrm{ph,crit}}$.
We begin our discussion by addressing the distinction between the
real and virtual phonon contribution. This we do partly due to 
somewhat confusing use of these terms in the existing literature. 

\subsection{Real and virtual phonons}

The phonon mediated effective electron--electron interaction 
${\cal D}_{21}(q,\omega)$, obtained here from diagrammatic perturbation
theory, can also be derived (with a little more work!) 
using elementary time-dependent perturbation theory starting from an expression
of the form\cite{landaulif}
\begin{equation}
{\cal D}_{21}(q,\omega) = \sum_{I} 
\frac{\langle i | \hat{H}_{e-{\rm ph}} | I\rangle 
\langle I | \hat{H}_{e-{\rm ph}} | f\rangle}
{E_{i} - E_{I} + i\eta}.
\end{equation}
Here $|i\rangle$, $|f\rangle$, and $|I\rangle$ are the initial state,
the final state in which momentum has been transferred between the 
layers, and an intermediate states in which the momentum to be
transferred is carried by a non-equilibrium phonon.
The infinitesimal imaginary part, $i\eta$,
in the denominator enforces causality\cite{landaulif} and plays
a crucial role when intermediate and initial or final states
are close in energy.  In time-dependent perturbation 
theory, this form of effective interaction determines the transition rate 
between initial and final states when the perturbing term in the 
Hamiltonian, the electron-phonon interaction in the present case, 
does not directly couple initial and final states.  
We therefore expect the phonon mediated effective interaction to play the 
same role
in transport experiments as the interlayer Coulombic interaction which 
{\it does} have direct matrix elements between initial and final states.
The intermediate state need not conserve energy and, in the jargon
of time-dependent perturbation theory, is consequently referred to 
as a virtual state.  

The intermediate states are ones where a phonon of 
momentum ${\bf Q} = ({\bf q}, Q_z)$ has been created or
destroyed by an electron in layer 1 or 2. 
Summing over four intermediate states for
each $Q_z$ and performing thermal averages gives the following expression
for transition matrix elements with 2D wavevector transfer ${\bf q}$
\begin{equation}
{\cal D}_{21}(q,\omega)=\int\!\frac{dQ_z}{2\pi}
|M_{l}(Q)|^2F_{1}(Q_z)F_{2}(-Q_z)
\left[
\frac{1}{\omega-\omega_{\bf Q}+i\eta}-\frac{1}{\omega+\omega_{\bf Q}
-i\eta} \right].
\label{2nd-order-pt}
\end{equation}
Eq.~(\ref{effint}) reduces to this form for 
$\ell_{\rm ph}\rightarrow\infty$ except that 
we have, for simplicity, retained only the longitudinal phonons in
the present discussion.  

Recalling that $\hbar\omega$ is the energy transferred between the
layers and $\hbar\omega_{\bf Q}$ is the energy of the intermediate
phonon, when the denominator vanishes in Eq.\ (\ref{2nd-order-pt})
energy is conserved in the intermediate state.
Therefore, the real (imaginary) part of the term in the square parenthesis
gives the virtual (real) phonon contribution.
In analogy, we define the {\em virtual} phonon exchange contribution
to the effective interaction as the contributions from 
${\rm Re}D$ in Eq.\ (\ref{effint}), and the {\em real} phonon
exchange contribution as that from ${\rm Im}D$. 
Ignoring the anisotropy factors in the phonon matrix
elements for the deformation potential, it turns out that in
the $\ell_{\rm ph}\rightarrow \infty$ limit, the 
entire $\omega < c_l q$ contribution to ${\cal D}_{21}$ is ``virtual''
and the entire $\omega > c_l q$ contribution is ``real.''
In general, the division between the contributions are not
so clear cut; at a given $\omega$ and $q$ 
both virtual and real contributions could exist simultaneously. 

In a semiclassical transport theory, real phonon processes result in
a non-equilibrium distribution  
of phonons in a coupled electron--phonon Boltzmann equation.  
In Appendix \ref{boltzmannreal} we derive an expression 
for the real phonon exchange contribution to the drag resistance
using such a coupled Boltzmann equation approach 
and explicitly demonstrate its equivalence to the 
purely real phonon contribution of ${\cal D}_{21}$ to the drag.
The appearance here of both virtual and real phonon contributions 
to the transresistivity in a single Feynman diagram 
is reminiscent of the appearance of both contributions to the 
quasiparticle scattering rate\cite{migdal,holstein,allen}
in the phonon exchange contribution
to the electron self-energy of a 3D electron-phonon system.

\subsection{Reciprocal space calculation}

In the following subsections we elucidate the physics of the drag for short 
phonon mean free paths.
When $\ell_{\rm ph} \ll \ell_{\rm ph,crit}$, 
we can set the ${\cal D}_{ij} $ factors in the expression $\epsilon$
to zero.  The screened interlayer interaction is then simply
\begin{equation}\label{old}
W_{21}(q,\omega) \simeq  \frac{{\cal D}_{21}(q,\omega)}
{(1 - U_{11}(q)\chi(q,\omega))^2}
\simeq \frac{{\cal D}_{21}(q,\omega)}{\epsilon_{0}^2},
\end{equation}
where we have taken the static limit of $\chi$ 
appropriate for temperatures comparable to $T_{BG}$. 
The transresistivity is then given by
\begin{equation}\label{omintapp}
\rho_{21,l}\simeq\frac{-\hbar^2}{8 \pi^2 e^2n_{1}n_{2}k_{B}T}
\int_{0}^{\infty} dq\; q^3\  
\frac{{\rm Im}\chi_{1}(q,c_{l}q){\rm Im}\chi_{2}(q,c_{l}q)}
{\sinh^2(\hbar c_{l}q/2k_{B}T)}\frac{1}{\nu_{0}^2\epsilon_{0}^4}
\int_{0}^{\infty}\! d\omega\ 
\left| \nu_{0}{\cal D}_{21}(q,\omega) \right|^2.
\end{equation}
We have factored terms which vary slowly with respect to $\omega$
out of the $\omega$ integration, since ${\cal D}_{21}(q,\omega)$ 
is relatively sharply peaked around $\omega = c_l q$.

{}From Eqs.\ (\ref{gD_{i}j}) and (\ref{B_ij}), $\nu_0 {\cal D}_{12}$
for $\omega \approx c_l q$ is given approximately by 
\begin{eqnarray}
&&\nu_{0}{\cal D}_{21}(q,\omega)
\simeq
\nonumber\\
&&\left\{
\begin{array}{ll}
-q^2 C_{\rm DP}\ f_{\rm cut}(\tilde{q})\;\exp(- d\tilde{q})/
(k_F \tilde{q})
& , q - \omega/c_{l} \gg \ell_{\rm ph}^{-1} \\
q^2 C_{\rm DP}\ f_{\rm cut}(\tilde{q})\;\exp(id \tilde{q}
-\omega d/2c_{l}\ell_{\rm ph}\tilde{q})
/(ik_F\tilde{q}) & , 
 \omega/c_l - q \gg \ell_{\rm ph}^{-1}
\end{array}
\right.
\label{qzint} 
\end{eqnarray}
where $\tilde{q} =|q^2 - (\omega/c_l)^2|^{1/2}$, and $f_{\rm cut}(\tilde{q})$
 is a function (dependent on the form factor) which cuts off at 
$\tilde{q} \sim L^{-1}$.

In the $ c_{l} q > \omega$ case, 
the effective interaction is dominated by virtual 
phonon exchange, whereas in the $ \omega > c_{l} q$ case, the 
integral is dominated by energy conserving intermediate 
states and the effective interaction is due to real phonon
exchange.  The analytic expression for the integral is
complicated in the small region where 
$|\omega - c_{l} q| < c_{l}/ \ell_{\rm ph}$, and we ignore it
in our treatment below.  The smaller $\ell_{\rm ph}$ is, 
the wider this region becomes and hence the expressions derived
below are not quantitatively valid for small $\ell_{\rm ph}$ (below 
$10^4\,{\rm nm}$ for typical parameters in GaAs); however, 
the expressions seem to exhibit the correct qualitative
behavior when compared to numerical calculations even for small
$\ell_{\rm ph}$.

Parameters of experimental systems studied to date satisfy the 
inequalities $\ell_{\rm ph}\gg d>L$ and $(2k_{F}L)^2\gg 1$.
It follows that for $\omega$ near $c_{l} q$ 
the integrand of the frequency integral in the drag resistivity 
expression proportional to $\tilde{q}^{-2}\approx 2q |q - \omega c_l|$.
The logarithmic divergence of the integral is cut off at small $\tilde{q}$
by the validity limits in Eq.\ (\ref{qzint}) 
and at large $\tilde{q}$ by the exponential 
suppression factors or cut-off functions which appear in Eq.~(\ref{qzint}). 
For the real phonon contribution we find 
\begin{eqnarray}
\int_{q c_l}^{\infty}\! d\omega\ |\nu_{0}
{\cal D}_{21}(q,\omega)|^2
&\simeq&
\frac{q^3 C_{\rm DP}^2 c_l}{k_F^2}
\int_{\tilde{q}^r_{{\rm min}}}^{\tilde{q}^r_{{\rm max}}} d\tilde{q}\  
\frac{\exp(-q d/\ell_{\rm ph} \tilde{q})}{\tilde{q}}\nonumber\\
&=& 
\frac{q^3C_{\rm DP}^2 c_l}{k_{F}^2}
\left[{\rm Ei}\left(-d\sqrt{\frac{2 q}{\ell_{\rm ph}}}\right)
- {\rm Ei}\left(-\frac{q d L}{\ell_{\rm ph}}\right)\right]
\label{Ei1}
\end{eqnarray}
where
\begin{eqnarray}
\tilde{q}^r_{{\rm min}} &=& (q/2\ell_{\rm ph})^{1/2};  \nonumber\\
\tilde{q}^r_{{\rm max}} &=& L^{-1},
\end{eqnarray}
and ${\rm Ei}(x) \equiv \int_{-\infty}^x dt\ \exp(t)/t$ is the
exponential integral, which has the limiting behavior
\begin{equation}
{\rm Ei}(-x)\sim\left\{
\begin{array}{ll}
\ln(x)     & ,   0 < x \ll 1;  \\
-\exp(-x)/x & ,  x \gg 1.
\end{array}\right.
\end{equation}
The upper cut-off $\tilde{q}_{{\rm max}}$ reflects the fact that it is 
impossible to excite phonons with a $z$-wavevector larger than $L^{-1}$. 
The $\tilde{q}$ integration in Eq.\ (\ref{Ei1}) can by physically
interpreted as follows.
The velocity component of a phonon in the $z$-direction 
is approximately $c_l \tilde{q}/q$, and hence the time taken to 
travel a distance $d$ in the $z$-direction is 
\begin{equation}
t_{\rm trans} = qd/c_l \tilde{q}
\label{transit-time}
\end{equation}
The exponential factor in Eq.\ (\ref{Ei1})
is $\exp(-t_{\rm trans} c_l/\ell_{\rm ph})$, 
which is the probability that a real phonon
emitted from one layer reaches the other. 

For the virtual phonon contribution we find 
\begin{eqnarray}\label{virres}
\int_{0}^{q c_l}\! d\omega\ |\nu_{0}{\cal D}_{21}(q,\omega)|^2
&\simeq&
\frac{q^3 C_{\rm DP}^2 c_l}{k_F^2}
\int_{q^v_{\rm min}}^{\infty} d\tilde{q}\ \frac{\exp(-2 d \tilde{q})}
{\tilde{q}}
\nonumber\\
&=&
-\frac{q^3 C_{\rm DP}^2 c_l}{k_{F}^2}\;
{\rm Ei}\left(-d\sqrt{\frac{2 q}{\ell_{\rm ph}}}\right).
\end{eqnarray}
where $q^v_{z,{\rm min}} = (q/2\ell_{\rm ph})^{1/2}$ is also given
by the validity limit in Eq.\ (\ref{qzint}).  The term in the 
exponent can be interpreted
as the probability that a virtual phonon emitted by one layer reaches the
other. The lifetime of the virtual state which is given by the
energy--time uncertainty relation 
$\Delta t \sim (q /c_l - \omega)^{-1} \sim \tilde{q}^2/c_l q$. 
The probability is $\exp(-t_{trans}/\Delta t)$,
where transit time is given in Eq.\ (\ref{transit-time}).
This gives the exponential factor and hence the distance
dependence in Eq.~(\ref{virres}).

Only the sum of the two contributions will be observed in experiments. 
Adding the two terms yields
\begin{eqnarray}\label{logres}
\int_{0}^{\infty}\! d\omega\ |\nu_{0} {\cal D}_{21}(q,\omega)|^2
&\approx&-\frac{q^3c_{l}C_{\rm DP}^2}{k_{F}^2}
{\rm Ei}\left(-\frac{q d L}{\ell_{\rm ph}}\right)
\nonumber\\
&\approx&
\frac{q^3c_{l}C_{\rm DP}^2}{k_{F}^2}\times\ 
{\left\{
\begin{array}{ll}
\ln\left(\frac{\ell_{\rm ph}}{qLd}\right),
&    
q L d/\ell_{\rm ph}  \ll 1;  \\
\ell_{\rm ph} \exp(-q L d/\ell_{\rm ph})/(q L d) \ \ ,
&
q L d/\ell_{\rm ph}  \gg 1.  \\
\end{array}\right.}
\end{eqnarray}
Inserting Eq.~(\ref{logres}) in Eq.~(\ref{omintapp}) we obtain  
\begin{equation}\label{simple}
\rho_{21}\approx\left(\frac{h}{e^2}\right) K_\lambda
\frac{{\rm Ei}(-2k_{F}Ld/\ell_{\rm ph})}{n_{1}n_{2}k_{B}T}
\int_{0}^{\infty}\! dq\left(\frac{q^\alpha}{\epsilon_{0}^4}\right)
\frac{{\rm Im}\chi_{1}(q,c_{\lambda}q){\rm Im}\chi_{2}(q,c_{\lambda}q)}
{\sinh^2(\hbar c_{\lambda}q/2k_{B}T)},
\end{equation}
where $K_\lambda$ was defined in Eq.~(\ref{longcst}) and again
$\alpha=6$ for $\lambda = l$ and $\alpha=2$ for $\lambda=t$.
The argument of the exponential integral has been replaced by the typical value
$\ell_{\rm ph}/2Lk_{F}d$ since it varies slowly over the large 
angle scattering region which dominates the integral ($k_F = {\rm
min}[k_{F,1},k_{F,2}]$).  Eq.\ (\ref{simple}) gives low $T$ power
laws for $\rho_{21}$ of $T^{10}$ for $\lambda=l$ and $T^6$ for
$\lambda=t$.

The distance dependence in the short phonon mean free path limit has
now been made explicit; as $d$ is increased, the transresistivity
falls logarithmically
when $\ell_{\rm ph}/2Lk_{F}d \gtrsim 1$ and exponentially when
$\ell_{\rm ph}/2Lk_{F}d \lesssim 1$.
The dependence of the
transresistivity upon temperature, electron density and the ratio of the 
electron densities $n_{1}/n_{2}$ is given by the remaining integral over 
wavevector which essentially expresses the phase-space available for 
large angle scattering as we shall discuss in Section~\ref{num}.

\subsection{Real space calculation}\label{picture}

In many physics problems, translational invariance and time-independence
makes reciprocal space calculations, like the one presented in the 
preceding section, simple and convenient.  It is often the case, 
however, that the corresponding real space calculations provide 
a useful alternate language for physical interpretation.  In the 
present case the frequency integral in Eq.~(\ref{omintapp}) is
equivalent to a integral over time
\begin{equation}
\int_{-\infty}^{\infty}\! d\omega 
|\nu_{0}{\cal D}_{21}(q,\omega)|^2=
2\pi\int_{-\infty}^{\infty}dt\ |\nu_{0}{\cal D}_{21}(q,t)|^2
\end{equation}
where 
\begin{equation}\label{texpr}
\nu_{0}{\cal D}_{21}(q,t)\simeq\frac{c_{l}qC_{\rm DP}}{2\pi k_{F}}
\int_{-\infty}^{\infty}dQ_z\ 
F_{1}(Q_z)F_{2}(-Q_z)D({\bf Q},t),
\end{equation}
and the real time, retarded, phonon propagator is given by
\begin{equation}
D({\bf Q},t)=-2\Theta(t)\sin(\omega_{{\bf Q}} t) e^{-t/2 \tau_{\rm ph}}.
\end{equation}
where $\tau_{\rm ph} = \ell_{\rm ph}/c_{l}$.
Since the important values of $Q_z$ are small compared to $q$, we can expand
\begin{equation}
\omega_{{\bf Q}}=c_{l}\sqrt{q^2+Q_z^2}\approx
c_{l}q+\frac{c_{l}}{2q}Q_z^2.
\end{equation}
Notice that the phonons have a quadratic dispersion in the $Q_z$-direction 
but a linear dispersion in the plane of the electron layers. 
The $Q_z$-integration in Eq.~(\ref{texpr}) can now be performed if we 
model the density profiles as Gaussians, so that 
$F_1(Q_z)F_2(-Q_z)=e^{-L^2 Q_z^2}
e^{-idQ_z}$.  We find that 
\begin{equation}\label{crude}
\int_{-\infty}^{\infty}\! d\omega\ |\nu_{0}{\cal D}_{21}
(q,\omega)|^2
\approx\frac{c_{l}^2q^2C_{\rm DP}^2}{Lk_{F}^2}
\int_{0}^{\infty}dt\ e^{- t/\tau_{\rm ph}} \frac{\exp(-\frac{1}{2}
\frac{d^2}{L^2+(\frac{c_{l}t}{2qL})^2})}
{\sqrt{L^2+(\frac{c_{l}t}{2qL})^2}}.
\end{equation}

This expression suggests a picture in real time for the phonon
mediated interaction.  The integrand of the time integral is recognized as 
a wavepacket which is centered around one well and which broadens and 
decays 
as time evolves. When the wavepacket is broad enough to reach the 
other well, a transfer of momentum (parallel to the layers) can
occur. 
The distance dependence of the interaction is 
determined by the time-dependence of the intensity of the phonon field 
disturbance at the other well.  
Since $d^2\gg L^2$ there will be no contributions to the integral 
in Eq.~(\ref{crude}) until $t\simeq t^{*}= qLd/c_{l}$
which corresponds to the time it takes the wavepacket to reach the 
second well. 
For $t>t^{*}$ the square of the amplitude of the wavepacket at the 
second well (the integrand in Eq.~(\ref{crude})) falls off as 
$1/t$ until it is eventually cut off at  
$t\simeq\tau_{\rm ph}$.   If $\tau_{\rm ph} < t^*$, then
only the exponentially small tail of the phonon field impinges
on the second well.
Defining $s\equiv c_{l}t/qLd$ the time integral,
Eq.~(\ref{crude}) can be written
\begin{equation}
\int_{-\infty}^{\infty}\! d\omega\ 
|\nu_{0}{\cal D}_{21}(q,\omega)|^2
\approx \frac{2C_{\rm DP}^2 c_l q^3}{k_F^2}
\int_{1}^{\infty}ds\ \frac{1}{s}
\exp\left(-qL\frac{d}{\ell_{\rm ph}}s\right)\exp\left(-\frac{2}{s^2}\right).
\end{equation}
This integral cannot be evaluated analytically, but as $s > 1$ 
the term $\exp(-1/2 s^2)$ can be approximated by unity. 
The resulting integral then yields Eq.\ (\ref{logres}).
Therefore, the $d$-dependence of $\rho_{21}$ can be interpreted 
as coming from the interplay of the layer separation and 
the range of the phonons which mediate the drag.

In the regime of short phonon mean free paths, the drag is 
mediated by damped phonons some of which spread over large distances
in the $z$-direction. Hence, in this regime the boundaries in the 
$z$-direction {\it will} limit the effective $\ell_{\rm ph}$. 
When boundary scattering dominates, a complete theory would require a 
realistic description of phonons scattering off the sample boundaries.
This is in contrast to the limit $\ell_{\rm ph}\gg\ell_{\rm ph,crit}$
where the drag is dominated by a coupled mode which is localized on
the length scale of $d_B$ and boundaries beyond this range are irrelevant.

\section{Numerical results}\label{num}

The results presented in this section were obtained 
by numerical evaluation of the transresistivity formula
[Eq.(\ref{rho}] using the complete  
expressions for both the effective interaction 
[Eq.(\ref{effint})] and the screening function
[Eq.(\ref{eps})]. Although these numerical calculations
are free of some of the approximations used in the 
preceding sections in order to obtain transparent analytic
results, we do not expect them to be exact.
In particular, corrections to the random phase 
approximation for screening must have some quantitative
importance and are difficult to estimate reliably.

Except where noted, the layer density was chosen to have the typical 
value $n_{1}=1.5\times10^{11}$ cm$^{-2}$ which 
yields a Fermi temperature of $T_{F} = \varepsilon_{F}/k_{B} 
\simeq 60$~K.   The GaAs material parameters are 
taken from the literature and are given in 
Sec.\ \ref{approx-anal}.\cite{stor90} 

\subsection{Temperature dependence}\label{tempdep}

As mentioned in the introduction, the transresistivity increases
roughly as $T^2$ at low temperatures if the effective interaction
between the layers is frequency independent.  Deviations from this behavior 
are indicative of retarded effective interactions.  
In particular, phonon-mediated contribution to 
$\rho_{21}$ grows approximately as $T^{10}$ for the deformation
potential and $T^6$ for piezoelectric
contribution.  Both cross over to a linear 
dependence on $T$ at 
approximately their respective Bloch-Gr\"{u}neisen temperature.  
In Fig.\ \ref{tfig}, we plot $\rho_{21}/T^2$ as a function of $T$.
The shape of this curve is a result of an effective interaction 
which, for each $q$, is sharply peaked as a function of frequency
around $\omega = c_l q$. 
This peak in the frequency integral, which produces the dominant 
contribution to the phonon-mediated drag, is cut off exponentially
by the thermal phase space factor when $\hbar c_{l}q > 2k_{B}T$.
This implies that only phonons with energies less than the 
temperature can participate in the drag.  The crossover from $T^5$ or
$T^6$ (where the piezoelectric coupling dominates)
to linear $T$ dependence of $\rho_{21}$ (hence the peak in 
$\rho_{21}/T^2$) occurs when the $2 k_F$
phonons, which are responsible for most of the momentum transfer 
in the systems, can be excited.  
The peak is expected to occur near 
the temperature scale $T_{{\rm BG}}=\hbar 2 c_{l}k_{F}/k_{B}$ 
which is 7.8 K for the parameters used in Fig.\ \ref{tfig}.  
It in fact occurs at around $T_{\rm peak}\approx 2.5\,{\rm K}
\approx T_{\rm BG}/3$.  For $T \gtrsim T_{\rm BG}$ the contribution
of the deformation potential phonons is approximately 50 to 100
times greater than that of the piezoelectric phonons for the
parameters used in the figure.

We also plot the results of Eq.~(\ref{long-mfp-approx}) and the
experimental results from Ref.\ \onlinecite{gram-phon}.
One can see that the
long mean free path approximation, and small layer separation, 
Eq.\ (\ref{long-mfp-approx}) underestimates the contribution of the mode.  
An investigation of the integrand indicates that this is due to a rather long 
non-Lorenzian tail in the $\omega < c_l q$ regime which can
be seen in the solid curve of Fig.~\ref{intgrdfig} discussed later.

Comparison of the magnitude of the experimental points and the
theoretical curves seems to indicate that the effects of the
collective mode have been observed.  However, as we caution in
Sec.\ \ref{sub:mfp_dep}, this should not be construed as definitive
proof of the existence of the mode.

\subsection{Density dependence}

{}From Eqs.~ (\ref{long-mfp-approx}) and (\ref{simple}) 
the transresistivity at matched densities 
is proportional to $1/n^2$ times an integral whose value 
reflects primarily the allowed phase space volume for exchange of phonons 
with large planar momentum and small perpendicular momentum.
The maximum planar momentum transfer, 
$q=2k_{F}$, is proportional to the square root of the density. 
At temperatures larger than $T_{{\rm peak}}$, the phase-space 
integral should therefore increase with density reflecting the larger 
possible momentum transfers. 
At temperatures well below $T_{{\rm peak}}$, on the other hand, 
phonons at $q=2k_{F}$ cannot be exchanged, and a higher density cannot 
be exploited.  Consequently, in the case where the density is changed
at a fixed temperature the following behavior should be observed.  For low
densities, the increase in density $n$ permits larger momentum transfers,
thus increasing the transresistivity.  At some $n$, a further increase
in does no good as the $2 k_F$ wavevectors cannot be excited, and this
leads to an overall decrease in decrease in $\rho_{21}$.  
The resulting curve for $\rho_{21}$ has a bump or maximum as a function
of density.  The position of the bump or maximum increases with
increasing temperature, for the reasons stated above.
Fig.\ref{nfig} shows the
density dependence of the transresistivity for four different 
temperatures, $T=1,2,3$ and $4\,$K and $\ell_{\rm ph}=1\,\mu{\rm m}$
(short mean free path regime) and $1\,$mm (coupled-mode dominated
regime).  

While the general features discussed above are seen in both regimes, 
Fig.~\ref{nfig} still shows a qualitative difference between the two.
The reason is twofold. A comparison of Eqs. (\ref{long-mfp-approx}) 
and (\ref{simple}) reveals that there is an extra factor of 
${\rm Im}[\chi]$ in the integrand of Eq.\ (\ref{simple}).
Since ${\rm Im}[\chi]$ decreases with increasing $k_F$, 
$\rho_{21}$ falls faster with increasing density in the short mean 
free path regime than in the coupled-mode regime. More importantly,
$\ell_{\rm ph,crit}$ is strongly dependent on density:
$\ell_{\rm ph,crit}\propto (2k_F+q_{\rm TF})^2/k_F^5$. Hence, 
increasing the density can make the coupled mode more and
more dominant and lead to an increase of $\rho_{21}$. As seen in 
Fig.~\ref{nfig} the difference in density dependence could provide 
a clue regarding which of the two regimes prevails in a particular 
sample.

\subsection{Density-ratio dependence}

The transresistivity is strongly dependent on electron-density ratio,
$n_{1}/n_{2}$.  For temperatures well below the Fermi temperature,
the maximum planar momentum transfer is $q=2k_{F,{\rm min}}$ where 
$k_{F,{\rm min}}$ is the Fermi wavevector of the electron gas 
with the lowest density. Decreasing one density will lead to a decrease
in $\rho_{21}$; a decrease also occurs if one density is increased because of 
the general $1/n_{1}n_{2}$ dependence.  The resulting peak at equal
densities, illustrated in Fig. \ref{rfig}, 
is independent of temperature and does not occur for 
the Coulombic drag mechanism (in the absence of any plasmon 
effects\cite{flen94,flen95b}). 
This difference has been used\cite{gram-phon}  
to separate Coulomb and phonon drag mechanisms experimentally 
when the layer separation is sufficiently small that the 
Coulomb mechanism is of comparable importance.

The calculations were done for both short and long mean free path
regimes.  However, the shape of the curves are so much alike 
that it is unlikely that a density-ratio measurement can be used
to differentiate these two regimes in an experiment.

\subsection{Mean free path dependence}
\label{sub:mfp_dep}

In the preceding section we have explained that there is a
critical $\ell_{\rm ph}$ 
beyond which a new collective mode appears which enhances the overall
transresistivity.   In Fig.\ \ref{lphfig},
we plot numerical results for 
the transresistivity as a function of the mean free path 
for $T = 3\,$K and $d = 50\,$nm.  
For $\ell_{\rm ph} < \ell_{\rm ph,crit}$, 
we are in the regime where the electron-phonon interaction
is separately screened in the two layers and $\rho_{21}$ 
increases logarithmically with $\ell_{\rm ph}$.
We see in Fig.\ \ref{lphfig}
when the mean free path exceeds $\ell_{\rm ph,crit}$ and 
the collective mode starts to emerge, the transresistivity increases
more rapidly before saturating at a value given roughly by the 
estimate given in Eq.~(\ref{long-mfp-approx}).
By splitting the contributions into real ($\omega > c_l q$),
virtual ($c_l q-\omega \gg \omega_0$) and coupled-mode
($\omega \approx \omega_0$), we show in the inset of Fig.\ \ref{lphfig}
that the increase in $\rho_{21}$ which occurs when $\ell_{\rm ph}$ 
exceeds $\ell_{\rm ph,crit}$ comes mainly from the coupled-mode contribution. 

In the low temperature regime where the drag experiments have been  
performed, $\ell_{\rm ph}$ is certainly sample dependent but 
should be temperature independent for a given sample.   
Extrapolating the logarithmic $\ell_{\rm ph}$ dependence 
({\it i.e.}, which neglects effects of collective screening) 
in Fig.\ \ref{lphfig} for the $d=50\,$nm case, we find that 
the value of $\ell_{\rm ph}$ required to fit measured 
values\cite{gram-phon} of $\rho_{21}$ ($\approx 12\,{\mathrm m}\Omega$)  
are (literally!) astronomical in magnitude.
This could be taken as evidence for the importance of 
cooperative screening and collective modes in present 
experimental systems. 
On the other hand, our numerical calculations are dependent
on random phase approximations estimates of Im$\chi$ which 
are likely to require substantial numerical revision\cite{cep} 
especially in the important region near $q=2k_{F}$.
These corrections will increase Im$\chi$ and could possibly boost
the theoretically calculated $\rho_{21}$ to 
experimental values without invoking the coupled electron--phonon
mode.  The random phase approximation for the screening 
function can also lead to quantitative discrepancy.  
Hence, we do not claim that there is incontrovertible experimental 
evidence for the existence of the coupled electron-phonon mode. 

\subsection{Layer separation dependence}

It is possible to prepare a series of double-quantum-well systems
which are substantially identical apart from the separation between
the 2D electron layers.\cite{gram-phon,gram-aps} 
Experiments on such a series of 
samples will give a more certain indication of the operative phonon 
exchange regime than experiments on a single sample.  

In Fig.\ \ref{dfig} we plot the transresistivity 
as a function of the separation between the wells, for various 
phonon mean free paths.   For $\ell_{\rm ph}$ small enough that 
the collective mode has not developed, $\rho_{21}$
exhibits a purely logarithmic dependence on $d$, until $d$ exceeds
$\ell_{\rm ph}/(2 k_F L)$, when it begins to fall exponentially.
For $\ell_{\rm ph} \gtrsim  \ell_{\rm ph,crit}$, the behavior is
more complex.  The collective mode develops and $\rho_{21}$ is 
considerably enhanced.  For small $d$, our numerical results suggest 
that $\rho_{21}$ also decreases
logarithmically with $d$, in contradiction with the
$d$-independence implied in Eq.\ (\ref{long-mfp-approx}).  
Fig.\ \ref{intgrdfig}, which shows the integrand of the 
transresistivity expression as a function of frequency at $q=2k_F$,
indicates that there is a substantial
non-Lorenzian tail for $\omega < c_l q-\omega_0$ which is
not taken into account in the analysis leading to Eq.\
(\ref{long-mfp-approx}). This tail contributes to the nearly logarithmic
$d$-dependence in the regime where the collective mode contributes
significantly.  

Another surprising result is that in this regime,
$\rho_{21}$ does not decrease monotonically as $d$ is increased.
Instead there is a peak when $d \approx \sqrt{\ell_{\rm ph,crit}/k_{F}}$.
Beyond this layer separation, the electron-phonon collective 
mode of the double quantum well begins to decouple into two weakly coupled 
modes centered around each well.
This is illustrated in Fig.\ \ref{intgrdfig}, where we compare the 
frequency integrands 
for $d < \sqrt{\ell_{\rm ph,crit}/k_{F}}$  and $d > \sqrt{\ell_{\rm
ph,crit}/k_{F}}$.  
As $d$ is increased further beyond $\sqrt{\ell_{\rm ph,crit}/k_{F}}$ 
the transresistivity eventually resumes its decline.

\section{Summary and Conclusions}\label{conclusion}

Phonon exchange contributes importantly to the frictional
drag resistance between nearby electron layers and 
is the dominant drag mechanism at large two-dimensional layer separations.
In this paper we have reported on a thorough theoretical
analysis of this drag mechanism.  We find that drag 
includes contributions due to exchange of both virtual and 
real phonons, and that novel coupled collective modes of the 2D electron
and 3D phonon systems can play a role depending on 
sample geometry and material parameters.  
We distinguish two regimes based on the relationship between the 
phonon mean free path and a crossover length scale 
$\ell_{\rm ph,crit}$ which is typically of the order of $0.2$ mm.
It is possible that the mean free path for high quality MBE grown 
heterostructures can exceed $\ell_{\rm ph,crit}$ at low temperatures.

In the short mean free path regime the dominant drag processes at 
momentum transfer $q$ have energy transfer just below $c_l q$ 
for virtual and just above $c_l q$ for real phonons.
The drag rate in this case decreases logarithmically with layer separation
$d$, until $d$ reaches $d_a = \ell_{\rm ph}/2k_{F}L \sim \ell_{ph}$.
The weak layer separation dependence comes nearly entirely from the 
virtual-phonon exchange contribution.  For $ d > d_a$, the 
virtual-phonon exchange contribution is small 
and the real-phonon exchange contribution decreases exponentially with 
layer separation.
The real-phonon exchange contribution to the drag 
is consistent with expectations based on the coupled 
Boltzmann equations for the electron and phonon systems,
and the exponential fall off for $ d > d_a$ can be 
understood in terms of the decay length for 
the disturbance of the steady state phonon system 
as the result of current flowing in one of the electron
layers.

For samples with phonon mean free paths larger than
$\ell_{\rm ph,crit}$ a new collective mode involving 
both electronic and lattice degrees of freedom emerges 
below the continuum of 3D phonon energies with 2D 
wavevectors $q$ near $2 k_{F}$.  The existence of this 
mode enhances the drag.  In this regime the drag also has 
a roughly logarithmic layer separation until $d$ reaches another crossover
length $d_{B} = (1+q_{\rm TF}/2k_{F})/16k_{F}C_{DP}$. For typical samples 
$d_{B} \sim 0.5\ \mu$m.  At this layer separation the 
collective mode separates into weakly coupled modes 
associated with the individual 2D layers.  Although the  
the drag has a complicated and non-monotonic 
layer separation in this regime, it does ultimately decline
with increasing $d$. These findings are summarized in Table (\ref{sumtable}).

There are, unfortunately, few existing experiments on the long distance 
dependence of the phonon mediated drag resistivity. More experimental
work will be needed to understand the range of possible behaviors and 
their dependence on system parameters. 
We note, however, that the preliminary experimental results by Gramila et
al.\cite{gram-aps} are consistent with a logarithmic $d$-dependence.

\acknowledgements 

This work was supported in part by the National Science Foundation
under grant DMR-9416906. MCB is supported by the Danish Research Academy.
The authors acknowledge helpful conversations
with Werner Dietche, Jim Eisenstein, Tom Gramila, Antti-Pekka Jauho, 
M.~Reizer, Holger Rubel, and Ned Wingreen.

\appendix

\section{Boltzmann equation derivation of the phonon drag}\label{boltzmannreal}

A sketch of the derivation of the contribution due to real emission of 
phonons is given here.  This Boltzmann equation approach has be 
used previously to estimate the real phonon contribution to the
drag.\cite{gram-phon,gram-upub}

Assume that the distribution function of the driving 
layer is a drifted Fermi Dirac.  Then, the deviation
function $\psi_{i}({\bf k})\equiv \delta f_{0,i}({\bf k})/
[(1-f_{0,i}({\bf k})) f_{0,i}({\bf k})]$ (where $f_{0,i}$ is the equilibrium
distribution function in layer $i$) for a driving field $e{\bf E}_1$
in the $x$-direction is
\begin{eqnarray}
\psi_{1}({\bf k}) = \frac{\tau e E_{1}}{k_{B} T} v_{x}({\bf k}).
\end{eqnarray}

The generation of nonequilibrium phonons is given by 
the electron--phonon coupling, and is
\begin{eqnarray}
\left(\frac{\partial N({\bf Q})}{\partial t}\right)_{\rm gen}
&=& -\frac{2\pi}{\hbar} \frac{2}{V} 
\sum_k |g_{1}(Q)|^2\ \delta(\varepsilon_{\bf k+q} - 
[\varepsilon_{\bf k} + \hbar \omega_{\bf Q}])\ 
f_{0,1}({\bf k})\; (1 - f_{0,1}({\bf k + q})\ N_{0}({\bf Q}) \times
\nonumber\\
&&\ \ \ \ \ \ [\psi_{1}({\bf k}) - \psi_{1}({\bf k + q}) + \Phi({\bf Q})]
\label{dNdt-gen} \\
\Phi({\bf Q}) &=& \delta N({\bf Q})/(N_{0}(Q) [1 + N_{0}(Q)]),
\end{eqnarray}
where $\delta N$ is the nonequilibrium distribution of phonons
and $N_{0}(Q) = [\exp(\hbar\omega_Q/k_{B} T) - 1]^{-1}$. 
The coupling constant $g_{1}({\bf Q})$ is given by 
\begin{equation}
g_{1}({\bf Q}) = |M({\bf Q})|^2\ |F_{1}(Q_z)|^2.
\end{equation}
Since the phonon coupling constant is small, we ignore the
$\Phi({\bf Q})$ term in Eq.\ (\ref{dNdt-gen}) because it is
higher order in $g$.

Then, one can write the phonon generation rate (using
the identity $f_{0}(\epsilon) (1- f_{0}(\epsilon+\omega))
= [f_{0}(\epsilon) - f_{0}(\epsilon + \omega)]\; [N_{0}(\omega) + 1]$),
\begin{equation}
\left(\frac{\partial N({\bf Q})}{\partial t}\right)_{\rm gen}
= -\frac{2}{L_z}\ |g({\bf Q})|^2\ N_{0}(Q)[1 + N_{0}(Q)]
\frac{q \tau e E_{1}}{m^* k_{B} T} {\rm Im}\chi(q,\omega_{\bf Q})
\end{equation}

The phonon Boltzmann equation is
\begin{equation}
v_z \frac{\partial \Phi({\bf Q},z)}{\partial z} = 
\delta(z)\, 2 |g({\bf Q})|^2 \frac{q_x \tau e E_{1}}{m^* k_{B} T}
(-{\rm Im}\chi(q,\omega_{\bf Q})) - \frac{\Phi({\bf Q},z)}{\tau_{\rm ph}},
\end{equation}
whose solution is
\begin{eqnarray}
\Phi({\bf Q},z) &=& 
-\frac{2 |g_1({\bf Q})|^2 q_x \tau e E_{1}}
{m^* k_{B} T\, |v_z({\bf Q})|} {\rm Im}\chi_{1}(q,\omega_Q) 
\exp\left(-\left|\frac{z}{v_z({\bf Q})\tau_{\rm ph}}\right|
\right)\ \theta(z Q_z). \nonumber\\
\end{eqnarray}

The electron-phonon collision term in layer 2 is
\begin{eqnarray}
\left(\frac{\partial f}{\partial t}\right)_{2}
&=& -\frac{2\pi}{\hbar} \frac{2}{V} 
\sum_{\bf Q} |g_{2}({\bf Q})|^2\ 
\Bigl\{-\delta(\varepsilon_{\bf k+q} - 
\varepsilon_{\bf k} + \hbar \omega_{\bf Q})\; 
[1 + N_{0}({\bf Q})]\;\Phi(-{\bf Q}) 
\nonumber\\
&&
+\delta(\varepsilon_{\bf k+q} - 
\varepsilon_{\bf k} - \hbar \omega_{\bf Q})\; 
N_{0}({\bf Q}) \Phi({\bf Q}) \Bigr\}.
\end{eqnarray}

Using $\Phi(-{\bf Q}) = -\Phi({\bf Q})$, 
the total momentum transfer to the second layer is
\begin{eqnarray}
\left(\frac{\partial {\bf p}}{\partial t}\right)_{2}
&=& 
2\pi \int \frac{d{\bf q}}{(2\pi)^{2}}
\frac{{\bf q}q_x\tau eE_{1}}{m^* k_{B}T}
\int_{0}^{\infty} \frac{dQ_{z}}{2\pi}\nonumber\\
&& \times  
\frac{2 |g_{1}({\bf Q})|^{2}  |g_{2}({\bf Q})|^{2}}{v_{z}({\bf Q})}
\frac{\exp{(-d/v_{z}({\bf Q})\tau_{\rm ph})}}
{4\sinh^{2} (\hbar\omega_{Q}/2 k_{B}T)}
{\rm Im}\chi_{1}(q,\omega_Q) {\rm Im}\chi_{2}(q,\omega_Q).
\end{eqnarray}
Given that $\omega_Q = c Q$, the transresistivity can be expressed 
in the following form:
\begin{eqnarray}
\rho_{21} &=& \frac{E_2}{J_1} = \frac{-(dp_{2,x}/dt) m^*}
{e^2 n_1 n_2 E_1 \tau}
\nonumber\\
&=& \frac{-1}{8 \pi^2 e^2 n_1 n_2 k_B T} \int_0^\infty dq\ 
q^3 \int_{0}^\infty d\omega\  
\frac{{\rm Im}\chi_{1}(q,\omega) {\rm Im}\chi_{2}(q,\omega)}
{\sinh^2(\hbar\omega/2 k_{B}T)}
\nonumber\\
&&\frac{\omega^2 
|F_{1}(\sqrt{(\omega/c_l)^2 - q^2})|^2 
|F_{2}(\sqrt{(\omega/c_l)^2 - q^2})|^2 
|M(q,\sqrt{(\omega/c)^2 - q^2})|^4}
{c_l^2 (\omega^2 - (q c_l)^2)}
\nonumber\\
&&\ \ \ \ 
\exp\left(-\frac{d}{ c_l\tau_{\rm ph} \sqrt{1 - c_l^2 q^2 /\omega^2}}\right).
\label{real-contr-final}
\end{eqnarray}

In comparison, the expression for the ``real'' contribution to 
the transresistivity is 
\begin{equation}
\rho_{21} 
= \frac{-\hbar^2}{8\pi^2 e^2 n_1 n_2 k_B T}
\int_0^\infty dq\ q^3 \int_0^\infty d\omega\    
[{\cal D}_{21,{\rm real}}]^2 
\frac{{\rm Im}\chi_{1}(q,\omega){\rm Im}\chi_{2}(q,\omega)}
{\sinh^2(\hbar\omega/2 k_{B}T)}
\label{real-contr-green}
\end{equation}
where 
\begin{eqnarray}
{\cal D}_{21,{\rm real}}({\bf q},\omega) &=&
\int_{-\infty}^\infty \frac{d q_z}{2\pi\hbar}\ |M(q,q_z)|^2
F_1(Q_z) F_2(-Q_z)\ {\rm Im}D({\bf Q},\omega)\nonumber\\
{\rm Im}D({\bf Q},\omega) &=&
\left[\frac{1/2\tau_{\rm ph}} {(\omega + c_l \sqrt{q^2 + q_z^2})^2 +
(1/2\tau_{\rm ph})^2}\right]
- \left[\frac{1/2\tau_{\rm ph}} {(\omega - c_l \sqrt{q^2 + q_z^2})^2 +
(1/2\tau_{\rm ph})^2}\right].
\end{eqnarray}
For small $1/\tau_{\rm ph}$ the poles of the Lorenzian are 
\begin{eqnarray}
q_z &=& \pm\sqrt{(\omega + i/(2c_l\tau))^2 - q^2 }
\nonumber\\
&\approx& \pm\sqrt{\omega^2/c_l^2 - q^2} \pm 
i\frac{\omega}{2 \tau c_l^2\sqrt{\omega^2/c_l^2 - q^2}}.
\end{eqnarray}
As they lie close to the real axis, the Lorenzian can in 
general be approximated by $\delta(\omega \pm c_l \sqrt{q^2 + q_z^2})$.
The imaginary part of the pole, however, does affect the 
$\exp(-iQ_z d)$ term which comes from 
from the difference in the phases of the form factors $F_1(Q_z)$ and 
$F_2(-Q_z)$.
The $Q_z$-integration results in insertion of the imaginary part of the 
pole into the exponent, yielding 
\begin{eqnarray} 
{\cal D}_{21,{\rm real}}({\bf q},\omega) 
&\approx& \frac{-1}{\hbar} \left|M\left(q,\sqrt{\omega^2/c_l^2 - q^2}\right)\right|^2 \  
F_1\left(\sqrt{\omega^2/c_l^2 - q^2}\right)
F_2\left(\sqrt{\omega^2/c_l^2 - q^2}\right)
\nonumber\\
&&\ \ \ \ \frac{\omega}{c_l^2\sqrt{\omega^2/c_l^2 - q^2}}
\exp
\left(-\frac{d}{2 c_l \tau_{\rm ph} 
\sqrt{1 - c_l^2 q^2 / \omega^2}}\right).
\label{D-real}
\end{eqnarray}
Substituting Eq.\ (\ref{D-real}) into Eq.\ (\ref{real-contr-green}),
gives Eq.\ (\ref{real-contr-final}). 


\begin{table}

\begin{tabular}{|c|c|c|c|c|}
$\ell_{\rm ph}$   &  \multicolumn{2}{c|}{$\ell_{\rm ph}\ll\ell_{\rm ph,crit}$}
  &  \multicolumn{2}{c|}{$\ell_{\rm ph}\gg\ell_{\rm ph,crit}$} \\ 
\hline
Physics  &  \multicolumn{2}{c|}{Damped phonons}  &   \multicolumn{2}{c|}
{Coupled {\em e--ph} mode}\\ \hline
$d$  &  $d\ll d_a$  &  $d\gg d_a$    &   $d\ll d_{B}$  &  $d\gg d_{B}$  
\\ \hline
Distance dependence  & $\ln(d_a/d)$  &  $\frac{d_a}{d}\exp(-d/d_a)$  &  
Weak (logarithmic)  &  Complicated\\
\end{tabular}
\caption{Table summarizing the distance behavior for the
collective mode regime and for the damped phonon cases.}
\label{sumtable}
\end{table}

\begin{figure}
\vspace{0.5cm}
\caption{
The Feynmann diagram corresponding to the correlation function (\ref{corr}).
The triangles (the function $\Delta$) in each layer are connected by
phonon propagators (wiggly lines). The dots represent the {\it e--ph} 
coupling and the dashed lines are external current operators.
The frequencies and 2-dimensional wavevectors ${\mathbf q}$ are conserved
in each vertex, whereas the perpendicular components, $Q_z$ and $Q'_z$,
are independently integrated over.
}
\label{tria}
\end{figure}

\begin{figure}
\vspace{0.5cm}
\caption{
The two-dimensional electron density of states times the
screened phonon mediated interaction,
$\nu_0|{\cal D}_{21}(2k_F,\omega)/\epsilon(2k_F,\omega)|$ as a
function of frequency near the longitudinal resonance
$\omega\sim 2c_lk_F$. The dotted, dashed, and solid lines are for
$\ell_{\rm ph}$=0.1 mm, 0.3 mm, and 1 mm, respectively. Other
parameters are $d$=500 \AA, $L$=200 \AA, and the density $n=1.5\times
10^{11}$ cm$^{-2}$.  The coupled electron--phonon mode 
occurs when $\nu_0\,{\rm Re}{\cal D}_{21}/{\rm Re}[\epsilon] \le -1/2$. 
[see Eq.\ (\ref{w-21approx})]. The inset indicates that for the
given parameters the coupled electron--phonon mode develops at 
$q=2k_F$ when $\ell_{\rm ph} \protect\gtrsim 0.5\,{\rm mm}$.
}
\label{wplot}
\end{figure}

\begin{figure}
\vspace{0.5cm}
\caption{
The scaled transresistivity $\rho_{12}/T^2$ as a function of temperature
for (in increasing order) $\ell_{\rm ph} = 10^4, 10^5, 10^6
10^7$ and $10^8\,$nm.
The parameters used are $n=1.5 \times 10^{11}\,{\rm cm}^{-2}$, 
$L = 200\,{\rm\AA}$ and $d = 500\,{\rm\AA}$ in GaAs. 
The dots are data points obtained from Ref.\ \protect\onlinecite{gram-phon},
and the dotted line is given by Eq.\ (\ref{long-mfp-approx}).
Inset: Log-log plots for the theoretical curves $\ell_{\rm ph} = 10^4$
to $10^8$, showing the crossover from $T^6$ to $T$ behavior.  The dotted
lines are for reference.
} 
\label{tfig}
\end{figure}

\begin{figure}
\vspace{0.5cm}
\caption{
The transresistivity as a function of matched densities $n$ 
for (a) $\ell_{\rm ph} = 1\,\mu{\rm m}$ and (b) $\ell_{\rm ph} = 
1\,{\rm mm}$.  
The temperatures are $T=1\,{\rm K}, 2\,{\rm K}, 3\,{\rm K}$ and 
$4\,{\rm K}$ for dash--triple-dotted, dash-dotted, dashed and
solid lines respectively.  Other parameters as in Fig.\ \ref{tfig}.
The ratio $\ell_{\rm ph}/\ell_{\rm ph,crit}$ varies in (b) from 
0.55 for $n=0.5\times 10^{11}$ cm$^{-2}$ to 14.6 for 
$n=2.5\times 10^{11}$ cm$^{-2}$.
}
\label{nfig}
\end{figure}

\begin{figure}
\vspace{0.5cm}
\caption{
The scaled transresistivity $\rho_{21} T^{-1} n_1/n_2$ as a function 
of the density ratio $n_1/n_2$ 
for (in increasing order) $T = 1, 2, 3, 4, 5, 6\,{\rm K}$, 
$n_2= 1.5\times 10^{11}\,{\rm cm}^{-2}$, $d =50\,{\rm nm}$,
(a) $\ell_{\rm ph} = 1\,\mu$m, and (b) $\ell_{\rm ph} = 10\,$cm.
Other parameters as in Fig. \protect\ref{nfig}.
}
\label{rfig}
\end{figure}

\begin{figure}
\vspace{0.5cm}
\caption{
The transresistivity as a function of the mean free path, for
$d = 50, 3000$ and $5\times 10^4\,{\rm nm}$ (solid, dotted and 
dashed lines, respectively).  The density is $1.5\times 10^{11}\,{\rm
cm}^{-2}$, and $T=3$ K.  Inset: 
Real phonon, virtual phonon and coupled-mode contributions (dotted,
dashed and solid lines respectively) for transresistivity as a function
of mean free path, for $d=50$ nm, $T=3$ K.
}
\label{lphfig}
\end{figure}

\begin{figure}
\vspace{0.5cm}
\caption{
The transresistivity as a function of the well-separation for (in
increasing order) $\ell_{\rm ph} = 10^3, 10^4, 10^5, 10^6, 10^7$ and
$10^8\,{\rm nm}$, at $T=3$ K, $n=1.5\times 10^{11}\,{\rm cm}^{-2}$,
$L=200\,{\rm\AA}$.
}
\label{dfig}
\end{figure}

\begin{figure}
\caption{
The integrand as a function of $\omega$ for $q = 2k_F$, $\ell_{\rm ph} =
\infty$, for $d=50\,{\rm nm}$ (solid), $d= 3\times 10^3\,{\rm nm}$
(dashed) and $d=5\times 10^4\,{\rm nm}$ (dotted). Other parameters as
in Fig.\ \protect\ref{dfig}.  As the well-separation
increases, the electron--phonon mode decouples into two weakly coupled
individual modes, leading to the twin-peaked structure at 
$d=3\times 10^3\,{\rm nm}$.  At large well separations, the collective modes
do not contribute to transresistivity, and all the momentum transfer
is due to the real phonons.
}
\label{intgrdfig}
\end{figure}


\begin{thebibliography}{10}
\bibitem{footnotephonons} In our calculation we
follow the common practice of treating the phonon
system as uniform.  There is actually a difference of
$\sim 5 \%$ between sound velocities of GaAs and 
${\rm Al_{x}Ga_{1-x}As}$ for typical values of $x$.  Although
we have not performed detailed calculations which respect this
mismatch, we do not believe that doing so would 
materially alter the conclusions reached in this paper.
\bibitem{squares}
The interaction Hamiltonian entering Eq.~(\ref{smatrix}) should be the sum of
  $\hat{H}_{e-{\rm ph}}$ (see Eq.~(\ref{intham})) and a Coulomb interaction
  Hamiltonian. The dominating term in the resulting formula for the
  transconductivity is of the form of Eq.~(\ref{tcon}) but with ${\cal D}_{21}$
  replaced by ${\cal D}_{21}+U_{21}$.
\end{thebibliography}
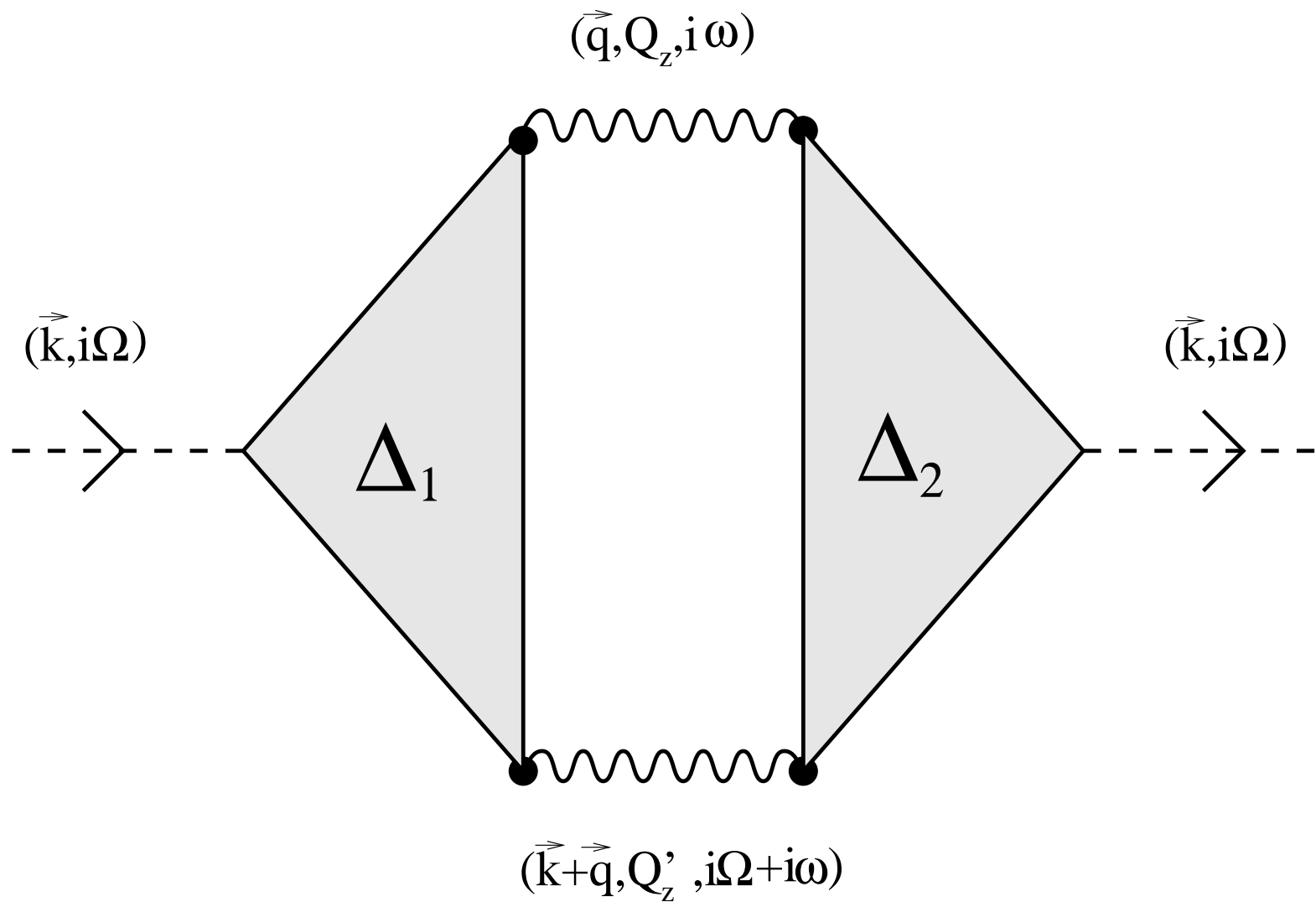
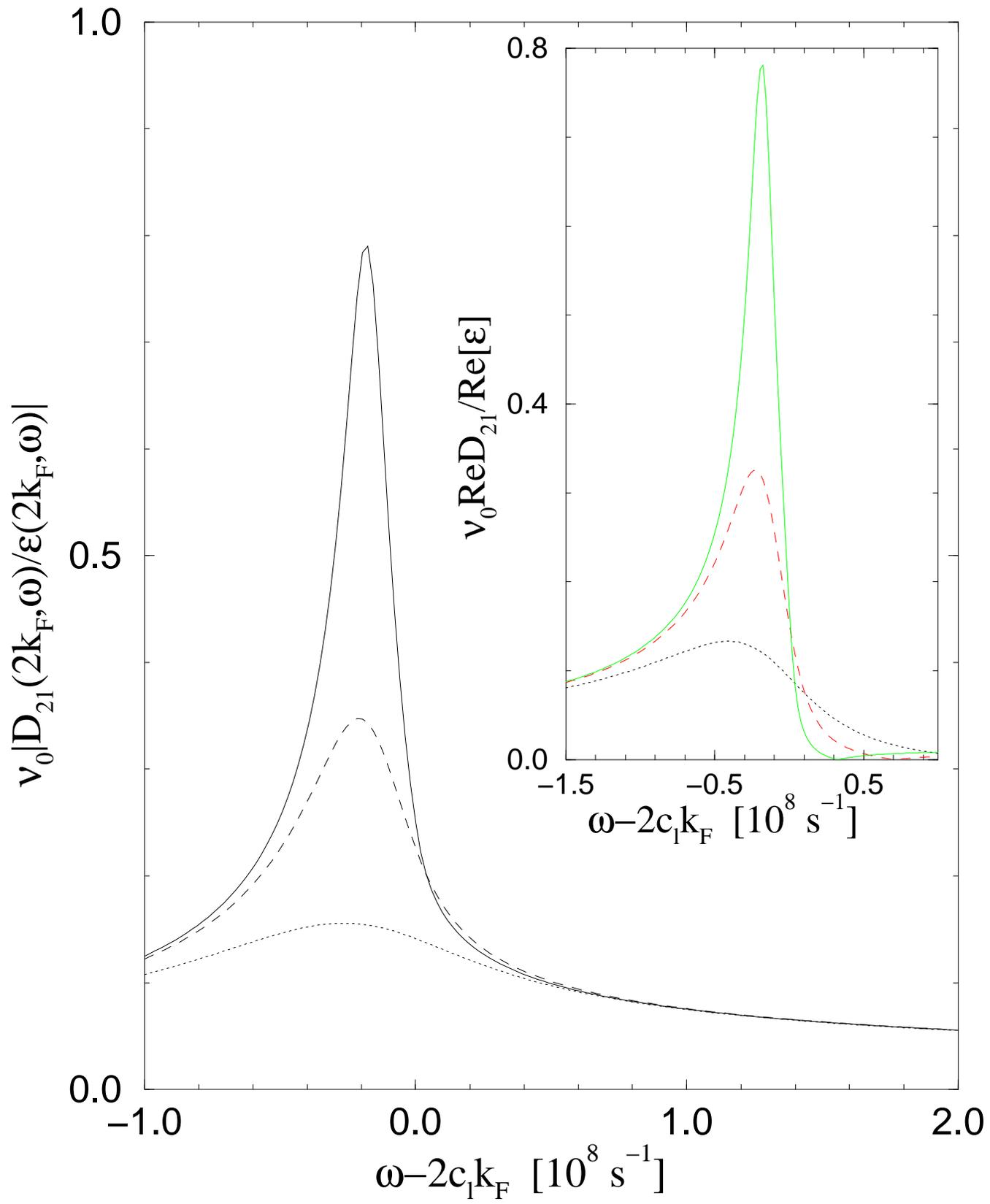
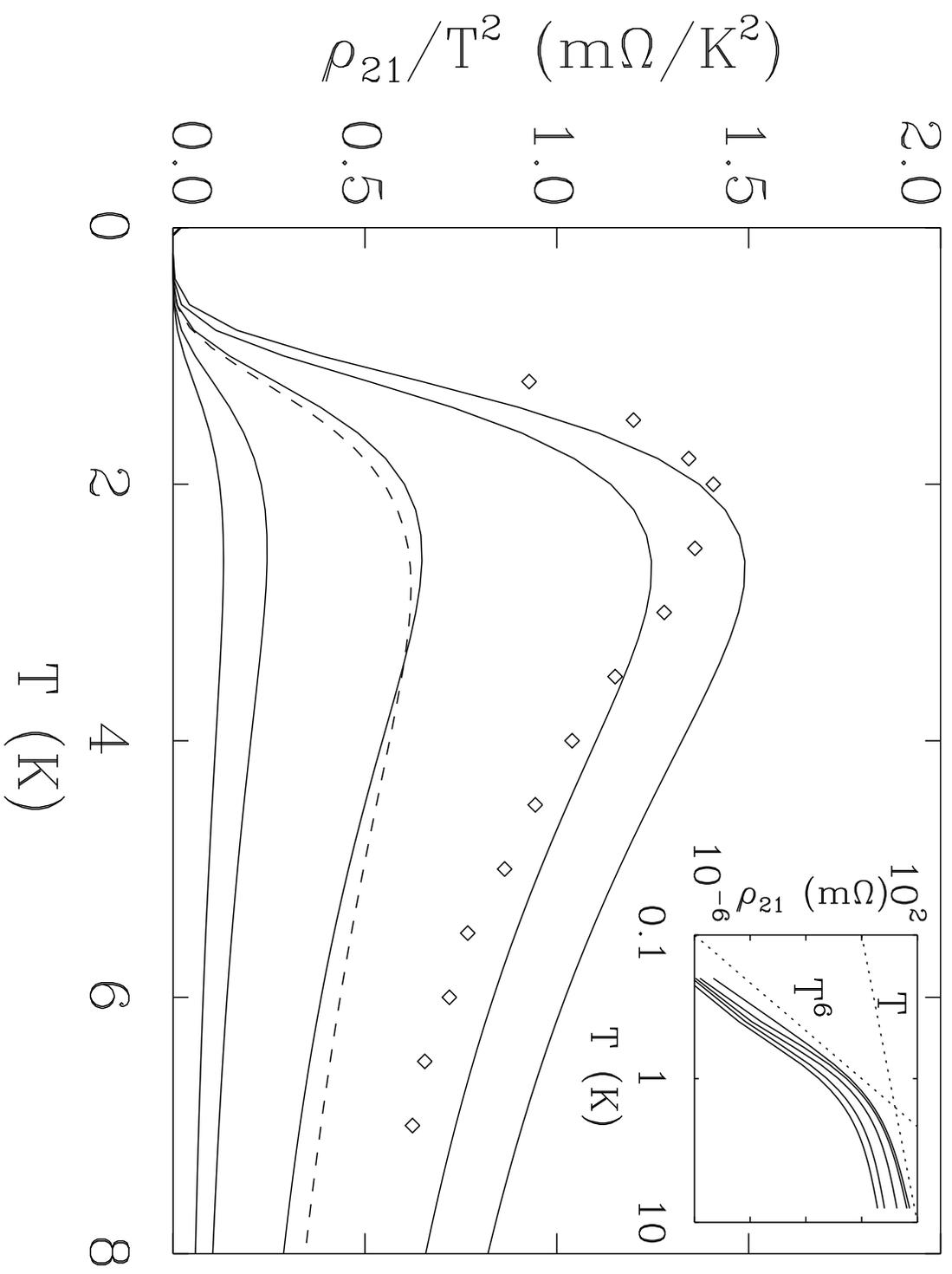
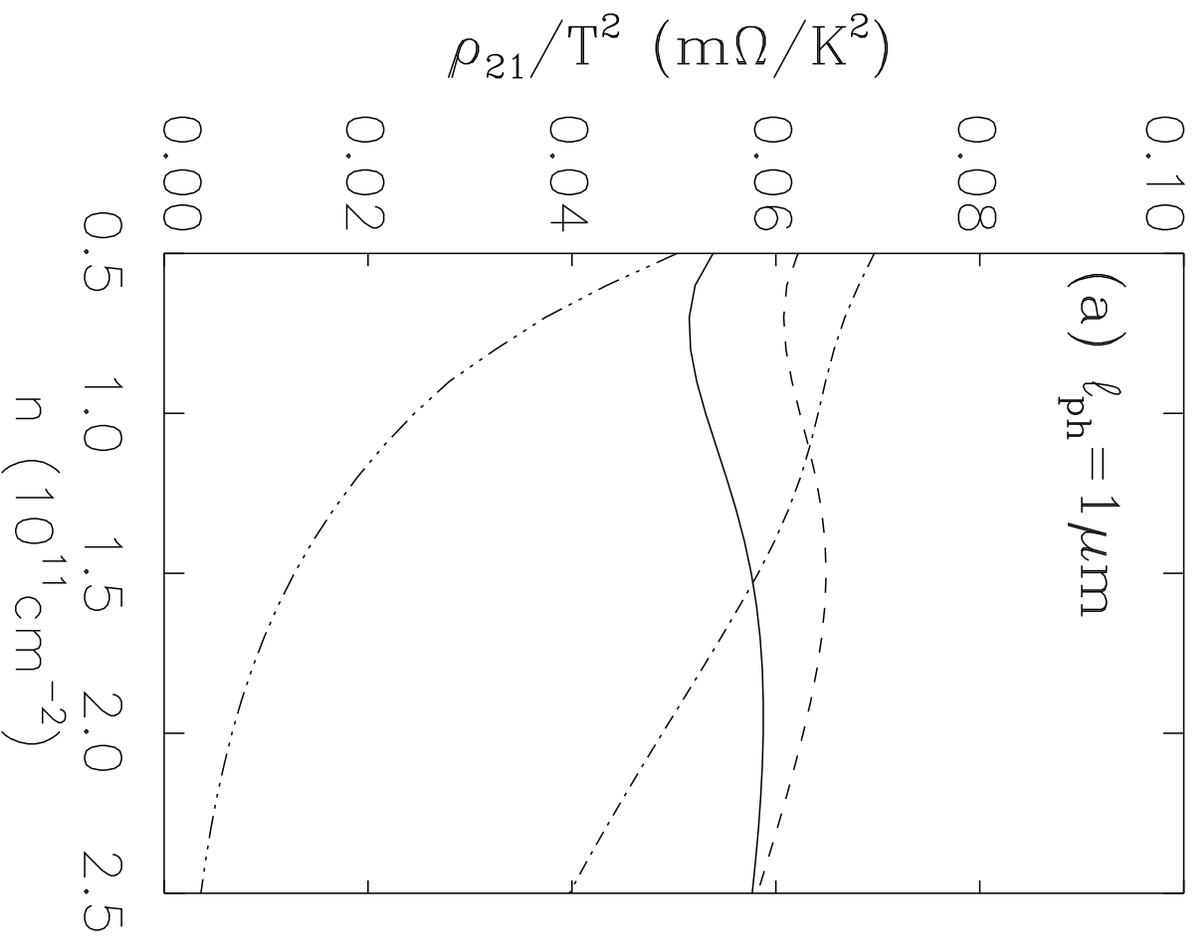
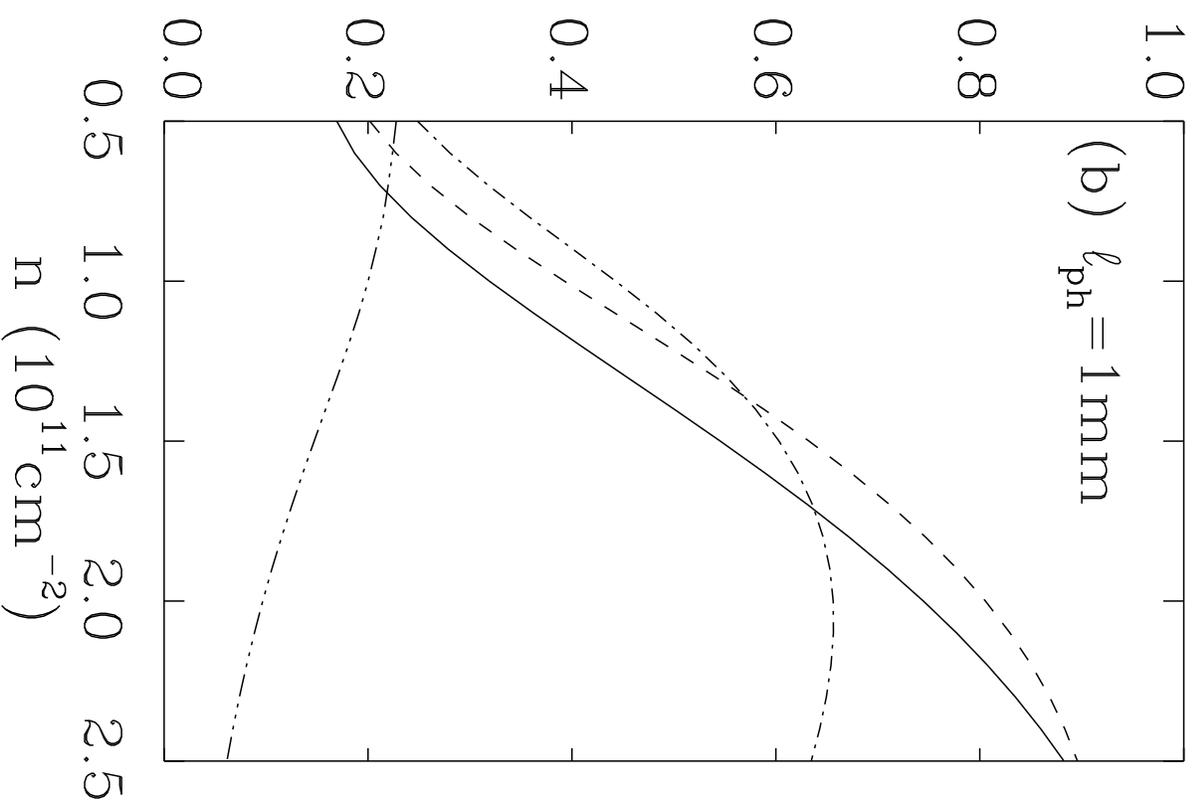
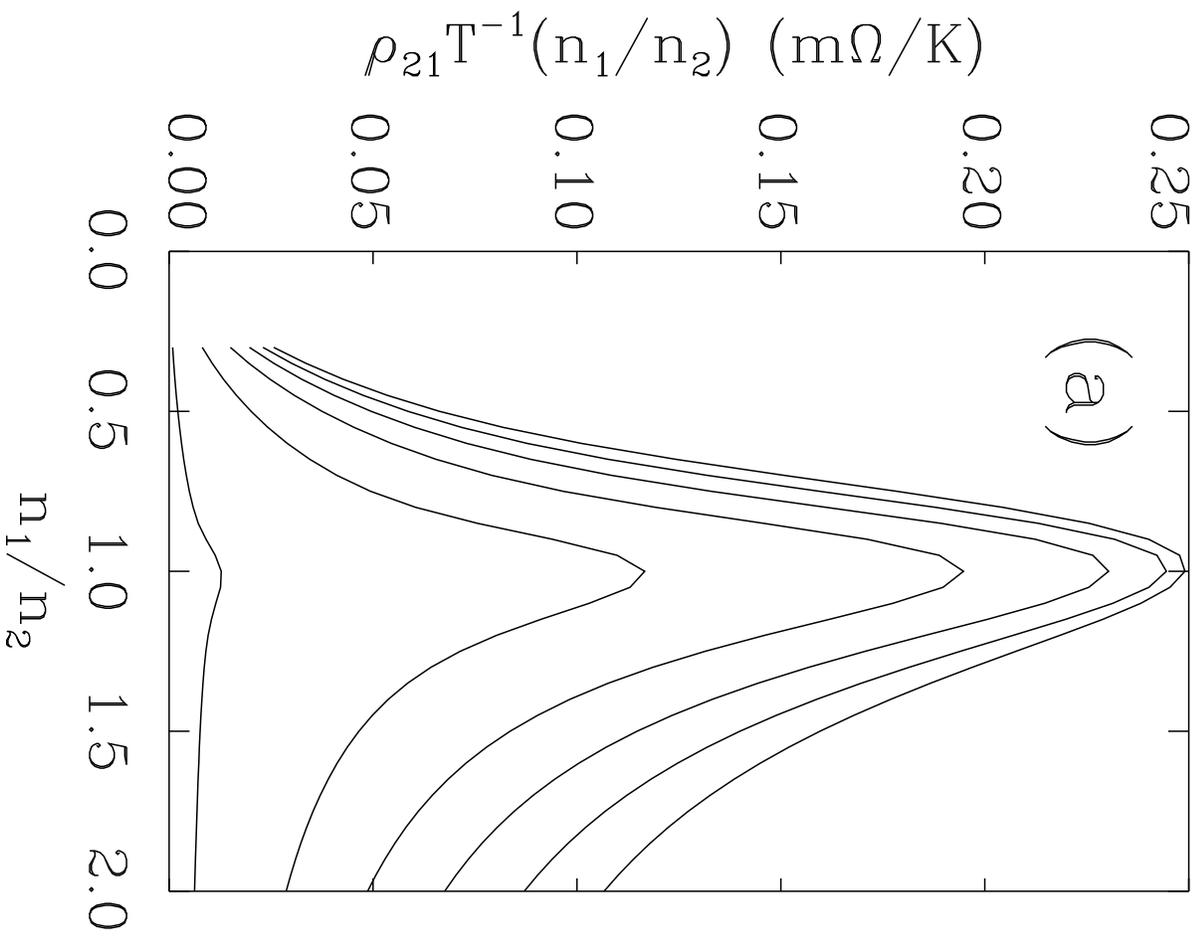
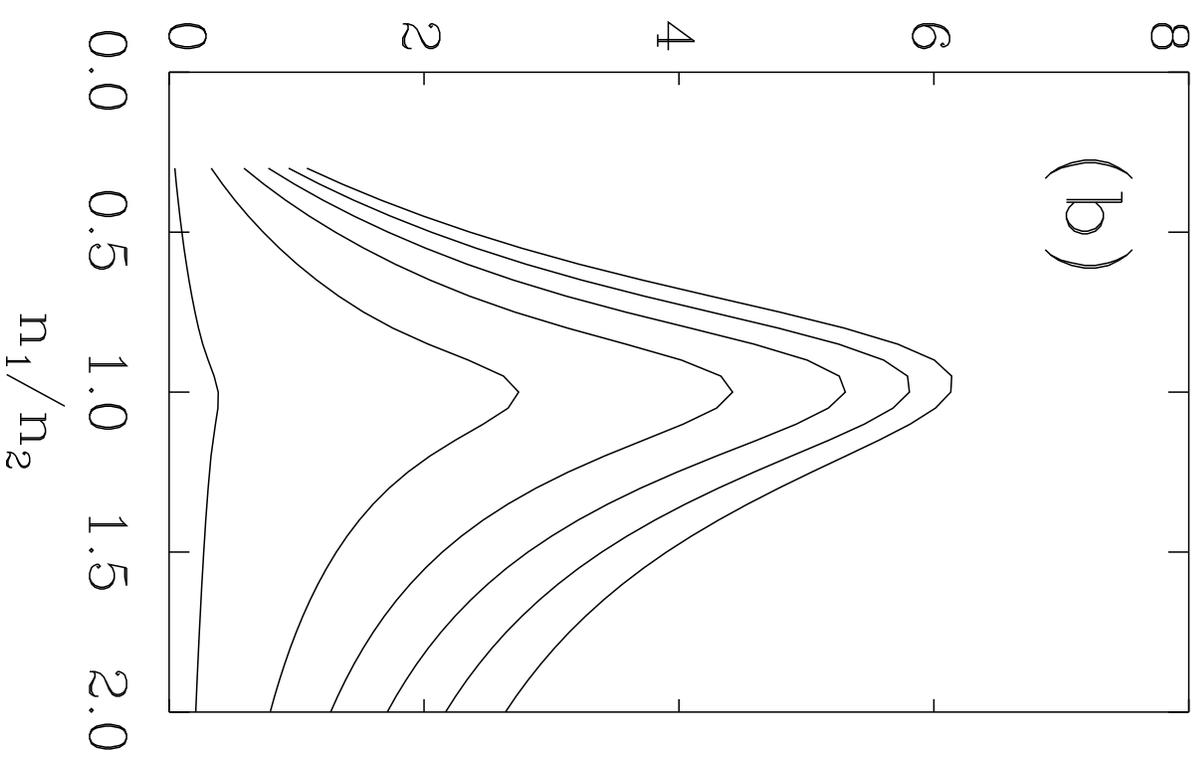
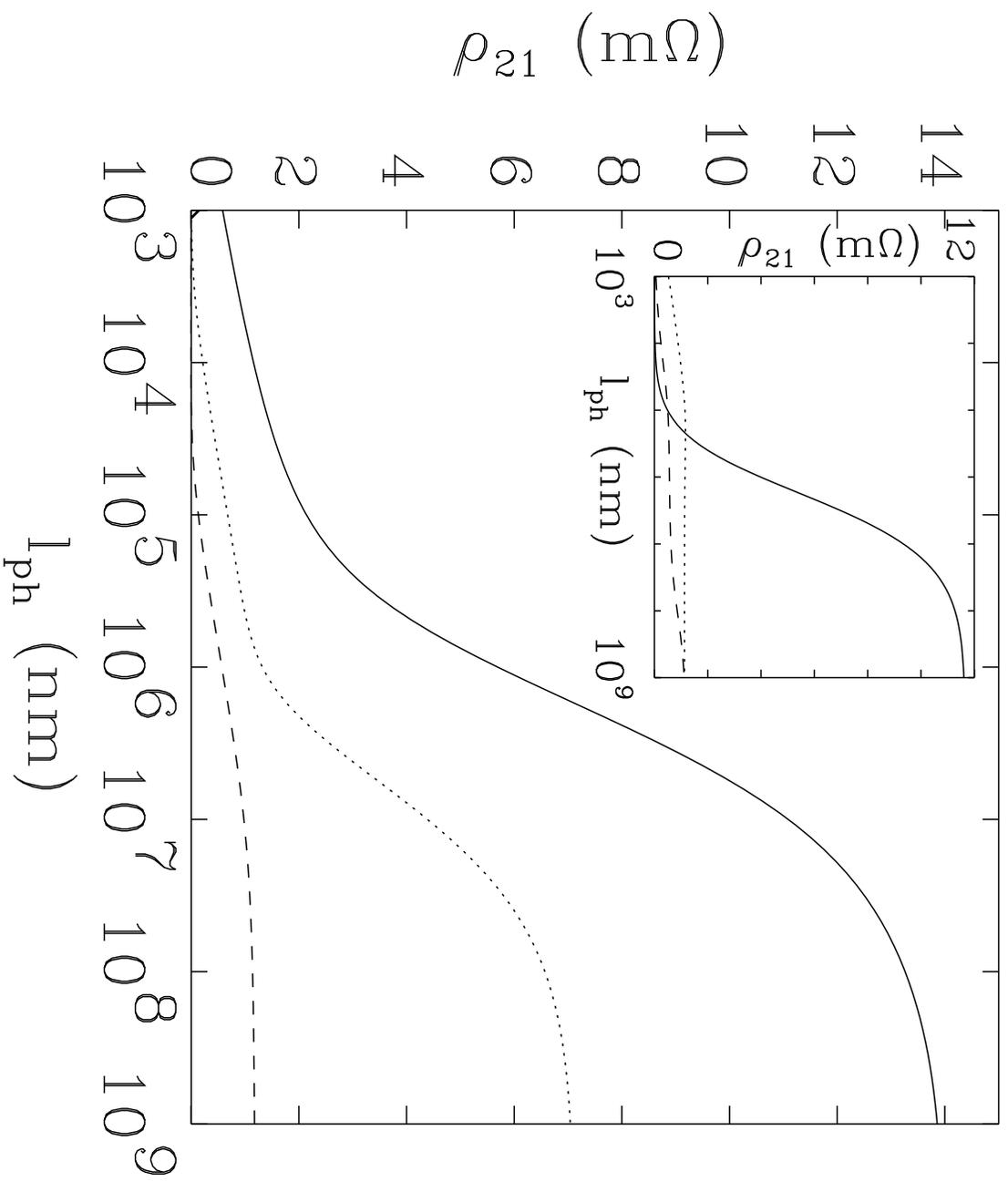
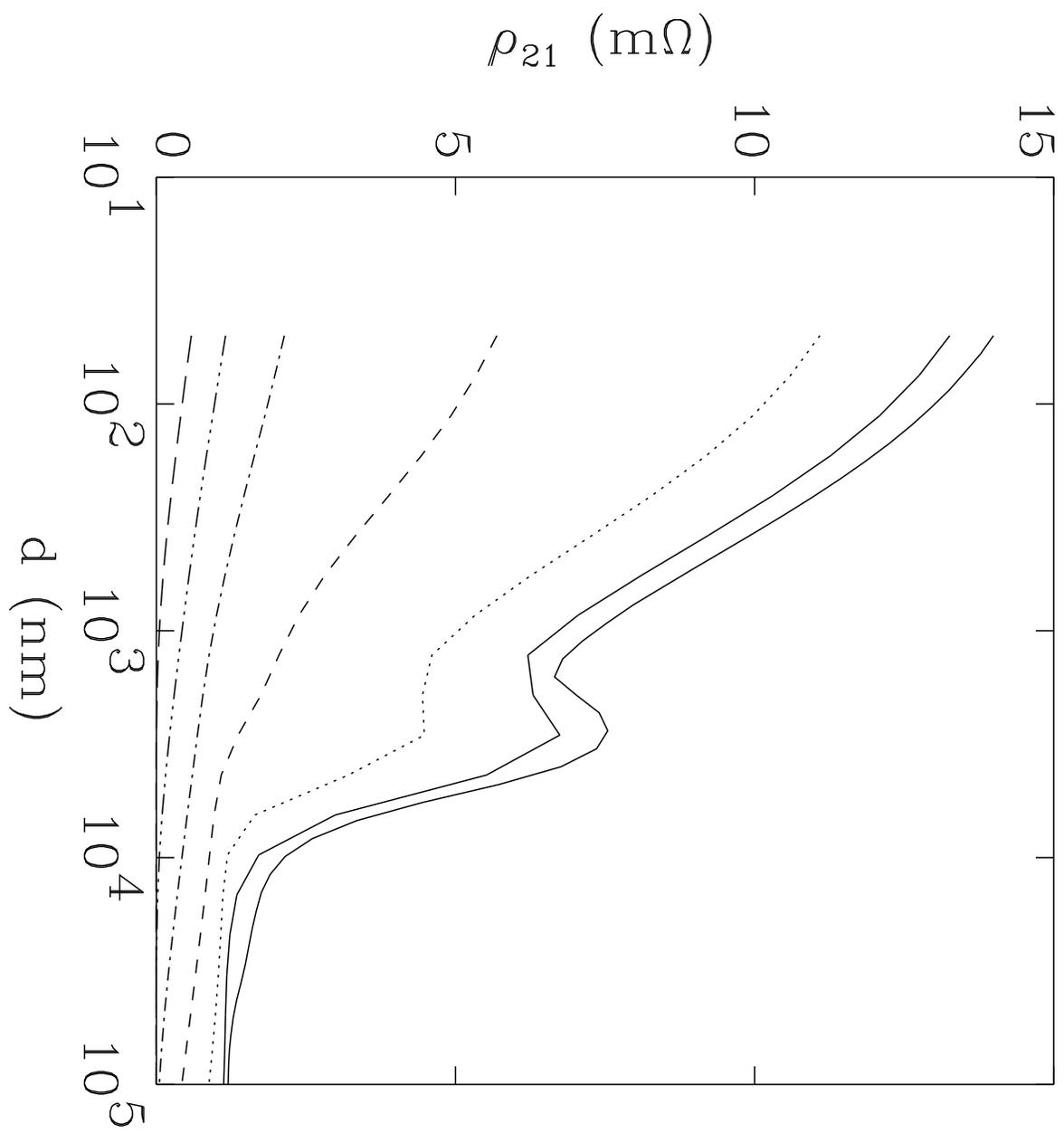
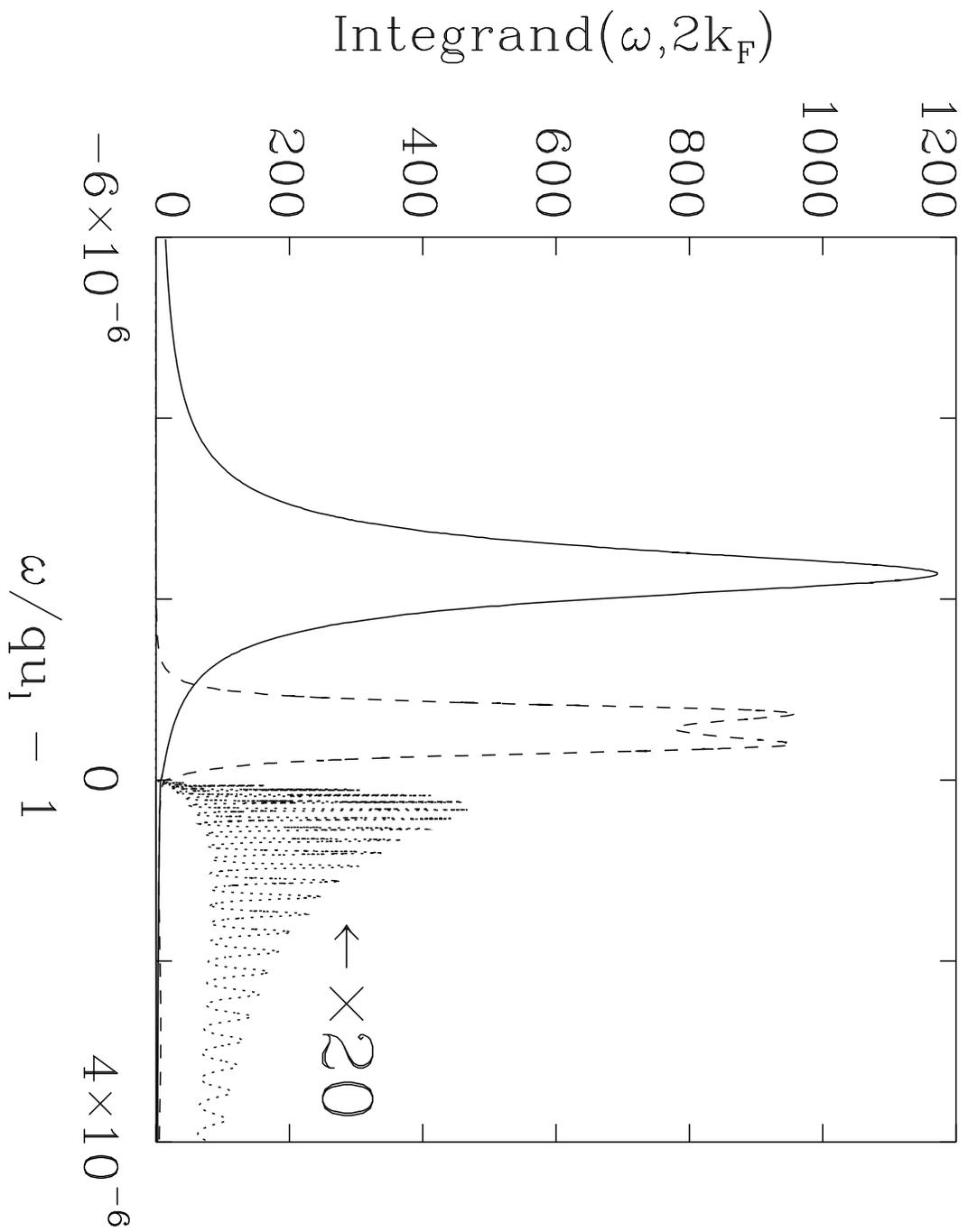
\end{document}